\documentclass[twocolumn,aps,prb,showpacs]{revtex4}
\usepackage{amsmath}
\usepackage{hyperref}
\usepackage{epsfig}
\usepackage{epsf}
\usepackage{array}
\usepackage{color}
\usepackage{graphicx}
\usepackage{epstopdf}
\usepackage{psfrag}
\usepackage{float}
\begin{document}
\bibliographystyle{apsrev}
\title{Physics of the Pseudogap in 8-site Cluster Dynamical Mean Field Theory: photoemission, Raman scattering, in-plane and  c-axis conductivity}
\author{Nan Lin, Emanuel Gull, and Andrew J. Millis}
\affiliation{Department of Physics, Columbia University, New York, New York 10027, USA}

\date{\today }

\begin{abstract}
Cluster dynamical mean field and maximum entropy analytical continuation methods are used to obtain theoretical estimates for the many-body density of states, electron self-energy,  in-plane and  c-axis optical conductivity and the $B_{1g}$ and $B_{2g}$ Raman scattering spectra of the two dimensional square lattice  Hubbard model at intermediate interaction strengths and carrier concentrations near half filling. The calculations are based on an 8-site cluster approximation which gives access to both the zone-diagonal and zone-face portions of the Fermi surface. At low dopings the  zone-face regions exhibit a `pseudogap', a suppression of  the many-body density of states for energies near the Fermi surface. The pseudogap magnitude is largest near half filling and decreases smoothly with doping, but as temperature is increased the gap fills in rather than closes. The calculated response functions bear a  strong qualitative resemblance to data taken in the pseudogap regime of high-$T_c$ cuprates, strongly suggesting that the intermediate coupling Hubbard model accounts for much of the exotic behavior observed in high-$T_c$ materials.  
\end{abstract}
\maketitle

\section{Introduction}
The `pseudogap', a temperature and carrier concentration dependent suppression of the many-body density of states of hole-doped high temperature copper-oxide superconductors which is visible at temperatures well above the superconducting transition temperature, is one of the enduring mysteries of the field.  The pseudogap was first inferred from  measurements of spin-lattice relaxation times \cite{Warren89} and Knight shifts;\cite{Alloul89} additional evidence rapidly accumulated from measurements of in-plane resistivity,\cite{Ito93} photoemission spectra,\cite{Loeser96,Ding96} the interplane (c-axis) conductivity,\cite{Homes93,Tajima97} Raman scattering \cite{Nemetschek97, Chen97} and tunneling \cite{Renner98} and measurements of the electronic density of states. On the other hand, the pseudogap does not lead to a significant suppression of low frequency spectral weight in the in-plane conductivity,\cite{Orenstein91} although structure in the conductivity and in scattering rates inferred from the conductivity has been attributed to the pseudogap.\cite{Basov06} The gap is most pronounced at low temperatures and low dopings. The data suggest that the gap magnitude decreases as doping increases, whereas with increasing temperature the gap magnitude does not decrease. Rather, the gap `fills in' as more and more states appear in the mid-gap region.  

Photoemission measurements \cite{Loeser96,Ding96,Damascelli03} indicate that the pseudogap is largest near the zone corner $(0,\pi)/(\pi,0)$ regions and vanishes at the zone-diagonal $(\pm \pi/2,\pm\pi/2)$ regions. This momentum-space structure of the pseudogap is one aspect of the more general phenomenon of momentum-space differentiation that characterizes the doping-dependent metal insulator transition in the high-$T_c$ materials. 

Schmalian, Pines and Stokjovic \cite{Schmalian98} argued that one should distinguish between a `weak' and `strong'  pseudogap. The weak pseudogap terminology refers to  relatively low energy phenomena affecting states within a few tens of meV of the Fermi surface while the  `strong pseudogap' is a suppression of density of states which may exist over a relatively wide energy range. 

The physical interpretation of the pseudogap remains controversial. One possibility is that it is associated with an actual thermodynamic transition to a phase with a definite long ranged order. Possibilities which have been proposed include magnetic order, perhaps of  `stripe'  or `nematic' density wave form \cite{Kivelson98} and an orbital current phase.\cite{Varma06} Experimental evidence has appeared supporting each of these possibilities \cite{Tranquada95,Fauque06} but the interpretations remain controversial.\cite{Sonier09}

An alternative possibility is that the pseudogap is a consequence of long but not infinite ranged order of  spin density wave,\cite{Millis93,Vilk97,Schmalian98,Abanov01} superconducting \cite{Emery95,Wang02} or RVB \cite{Kotliar88,Lee92,Altshuler96} type. One loop \cite{Lee73} and more sophisticated \cite{Vilk97,Schmalian98,Abanov01} calculations provide a relation between quasi-long-ranged order and  pseudogaps, while slave-boson-based mean field methods \cite{Kotliar88,Lee92,Altshuler96} provide a different mechanism for (pseudo)gap formation. However, these methods are based on uncontrolled analytical approximations. Models of quasi one dimensional `ladder'  compounds can be studied in a controlled manner and are known to exhibit gapped phases with no long ranged order,\cite{Dagotto96} however despite intriguing qualitative similarities the relation of `ladder' calculations to the two dimensional physics relevant to the cuprates remains unclear. There is a clear need for studies based on  methods which are applicable in two dimensions and at intermediate to strong couplings and which are not based on a particular assumption about the type of relevant correlations. 

One such approach is provided by `cluster' dynamical mean field methods.\cite{Maier05} These techniques are based on a coarse discretization of the  momentum dependence of the electron self-energy but permit a numerically unbiased solution of the resulting model. Important early work showed that the cluster dynamical mean field approximation produced features reminiscent of the pseudogap including suppression of the density of states in the $(0,\pi)$ region of the zone \cite{Huscroft01} and `Fermi arcs' which are at least qualitatively consistent with photoemission experiments.\cite{Parcollet04,Civelli05,Kyung06,Stanescu06} Subsequently many cluster dynamical mean field theory (DMFT) studies of the pseudogap  have appeared.\cite{Macridin06,Zhang07,Civelli08,Gull08,Park08,Ferrero09,Ferrero09b,Civelli09,Liebsch09,Sakai09,Sordi10,Sakai10}

Studies to date are mainly of two sorts: large-cluster studies of one doping and temperature and more comprehensive studies of smaller (two and four site) clusters. An important large cluster study is the work of Macridin {\it et al.},\cite{Macridin06} which considered a 16 site cluster at $U=8t$ at a doping of $5\%$ and an inverse temperature $\beta=8/t$ and showed from analytical continuation of the electron spectral function that a $(0,\pi)$ pseudogap existed at this doping.   The smaller-cluster studies have been based on  two and four site clusters for which the computational burden is much less, permitting systematic examinations of a wider range of parameter space.\cite{Civelli05,Kyung06,Gull08,Park08,Ferrero09,Civelli09,Zhang07,Liebsch09,Sakai09,Sordi10,Sakai10} However these smaller clusters do not allow direct comparison of zone-diagonal and zone-face regions of momentum space. In addition, many of the more systematic small cluster based analyses employed interpolation schemes to construct a representation of the electron self-energy throughout the Brillouin zone. The reconstructions are  based on data corresponding to the $(0,0)$, $(\pi,\pi)$ and, for the four-site clusters, $(0,\pi)/(\pi,0)$ momentum points \cite{Civelli05,Kyung06,Stanescu06} and the physical interpretation is problematic especially because information related to the physically important $(\pm \pi/2,\pm \pi/2)$ momentum space regions is inferred from measurements at $(0,0)$, $(\pi,\pi)$ and $(0,\pi)/(\pi,0)$ where the physics is presumably different. However,  an interesting recent paper by Ferrero {\it et al.}\cite{Ferrero09b,Ferrero10} introduced a two-site cluster with a new momentum space partitioning which clearly separated nodal and antinodal regions.

The work reported here builds on recent studies of the $8$-site cluster \cite{Werner09,Gull09} shown in Fig.~\ref{tiling}. The larger size of this cluster provides a more refined momentum resolution and in particular gives independent access to the zone-face and zone-diagonal regions of the Fermi surface. However the size is small enough that studies of wide regions of parameter space are computationally feasible. Previous work has revealed that the doping-driven Mott transition is  momentum-space-selective, with a gap opening first  in the zone corner $(0,\pi)/(\pi,0)$ regions of the Brillouin zone while  the zone-diagonal ($\pm\pi/2,\pm\pi/2$) regions remain ungapped until the carrier concentration reaches half filling. The previous work focused on quantities defined directly on the Matsubara axis. In this paper we   use analytical continuation techniques to examine the consequences of the momentum-space-selective transitions for observables including the electron spectral function and self-energy, the interplane and in-plane conductivity, and the Raman scattering intensity.  Our results strongly suggest that even in the absence of long ranged or quasi-long-ranged order, the Hubbard model at intermediate couplings contains the essential physics of the `strong pseudogap'.

The rest of this paper is organized as follows. In Section \ref{Model} we present the model and the dynamical mean field and analytical continuation methods we use to solve it. In Section \ref{Spectral} we show results for the electron spectral function and in Section \ref{caxis} the interplane conductivity. Section \ref{Raman} presents our results for the Raman scattering intensity and Section \ref{conductivity} the  in-plane conductivity.   Section \ref{Selfenergy} shows the electron self-energy in the different momentum sectors, confirming the conjecture that the gap arises from an orbitally selective Mott transition and demonstrating that the model reproduces key aspects of the momentum selectivity in the approach to the Mott transition.  Section \ref{Conclusion} is a summary and conclusion.

\section{Model and Methods\label{Model}}

We study the Hubbard model on a two dimensional lattice. The model is conveniently written in a mixed momentum-space/position-space representation as

\begin{equation}
H=\sum_{k\sigma}\varepsilon_kc^\dagger_{k\sigma}c_{k\sigma} +U\sum_i n_{i\uparrow}n_{i\downarrow}-\mu\sum_{i\sigma}n_{i\sigma}.
\label{H}
\end{equation}

We take, as a reasonable representation of the band structure of high temperature superconductors,
\begin{equation}
\varepsilon_k=-2t\left(\cos k_x+\cos k_y\right)-4t{'}\cos k_x\cos k_y
\label{dispersion}
\end{equation}
with  $t{'}=-0.15t$.  For comparison to data we note that a value generally accepted for high-$T_c$ superconductors is $t \approx 0.35\ \text{eV}$ \cite{Andersen94} while values of $t'/t$ from $-0.05$ to $-0.3$ have been reported for different materials.

To solve the model we use the `dynamical cluster approximation' (DCA).\cite{Hettler98,Maier05} The method is based on tiling the Brillouin zone into $N$ equal-area non-overlapping tiles and approximating the electron self-energy $\Sigma(k,\omega)$ as a piecewise constant function which may take different values in the different tiles. Labeling the tiles by center momentum $K$  we have
\begin{equation}
\Sigma(k,\omega)=\Sigma_K(\omega); \hspace{0.2in} k\in K.
\label{Sigmadca}
\end{equation} 

\begin{figure}[tbp]
\centering
\includegraphics[width=0.6\columnwidth]{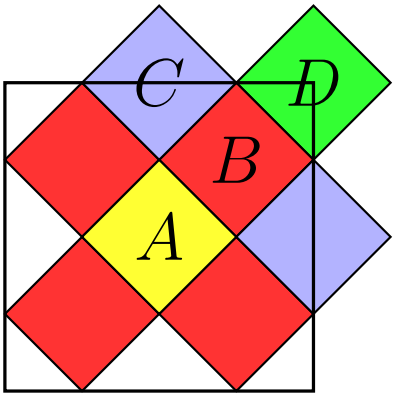}
\caption{Brillouin zone partitioning associated with the 8-site cluster.}
\label{tiling}
\end{figure}

The results we present were obtained using the  $8$-site momentum space partitioning shown in Fig.~\ref{tiling}  and the  `continuous-time auxiliary field' (CT-AUX) numerical method \cite{Gull08_ctaux} as discussed in more detail in  Ref.~\onlinecite{Werner09} and \onlinecite{Gull09}. Because the model is solved on the imaginary axis, an analytical continuation procedure is required to obtain real frequency information. Following Ref.~\onlinecite{Wang09} we continue the electron self-energies using the Maximum Entropy technique \cite{Jarrell96} and the `L-curve' method.\cite{Rabani02} The covariance matrix of the self-energies is approximately diagonal and the continuation of the obtained real frequency spectra back to the imaginary axis is in good agreement with the original data. Although uncertainties exist in the analytical continuation, our experience is that  the near Fermi-surface structures are reliable.

The $8$-site computation suffers from a fermionic sign problem which becomes worse with decreasing temperature and increasing doping and interaction strength. The need to scan a range of dopings and to obtain data of the precision required for  reliable analytical continuations limited us to temperatures $\geq t/20$ (corresponding roughly to $200\ \text{K}$ for $t=0.35\ \text{eV}$) and $U=7t$. The $U$-value is  smaller than the values  $U\sim 9t$ believed \cite{Comanac08} to describe high temperature superconductors, but is large enough that (within the approximation we use) the half filled state is a Mott insulator and hole-doping leads to a momentum-space-selective Mott transition.


Dynamical mean field methods involve a drastic simplification of the momentum dependence of the electron self-energy. As Eq.~(\ref{Sigmadca}) shows, the DCA method produces a piecewise constant self-energy which may be viewed as a discretization of the  continuous momentum dependence of the exact solution. In many previous dynamical mean field studies of the pseudogap  an interpolation process \cite{Civelli05,Stanescu06} was used to construct a self-energy with a continuous momentum dependence, which was then used to produce figures to be compared to experiment.  We prefer to avoid interpolations and instead work with analytical continuations of directly calculated quantities. We  present results for 
\begin{enumerate}
\item Sector-averaged electron spectral function
\begin{equation}
A_K(\omega)=-\frac{N_c}{\pi}\int_K\frac{d^2k}{(2\pi)^2}\text{Im}\left[\frac{1}{\omega-\varepsilon_k+\mu-\Sigma_K(\omega)}\right]
\label{Adef}
\end{equation}
with $N_c=8$ in our 8-site cluster.
\item Interplane or c-axis conductivity
\begin{eqnarray}
\sigma_c(\Omega)&=&2 \sum_K\int_K\frac{d^2k}{(2\pi)^2}\int\frac{d\omega}{\pi}\frac{f(\omega)-f(\omega+\Omega)}{\Omega}\times
\nonumber \\
&&t_\bot(k)\text{Im}G_K(k,\omega+\Omega)t_\bot(k)\text{Im}G_K(k,\omega)
\label{sigmacdef}
\end{eqnarray}
with $t_\perp(k)=t_0(\cos k_x-\cos k_y)^2$, $G_K(k,\omega)=\left[\omega-\varepsilon_k+\mu-\Sigma_K(\omega)\right]^{-1}$ and $f(\omega)$ the Fermi-Dirac distribution. We used $t_0=0.15\ \text{eV} \approx0.43t$ \cite{Andersen94} in the calculation.
\item Quasiparticle contribution to Raman scattering intensity
\begin{eqnarray}
\chi''(\Omega)&=&2\sum_K\int_K\frac{d^2k}{(2\pi)^2}\int\frac{d\omega}{\pi}[f(\omega)-f(\omega+\Omega)] \times
\nonumber \\
&&\gamma_{ab}\text{Im}G_K(k,\omega+\Omega)\gamma_{ab} \text{Im}G_K(k,\omega).
\label{ramandef}
\end{eqnarray}
Two geometries are of interest: $B_{1g}$, where the matrix element is $\gamma_{ab}=\frac{1}{2}(\partial^2\varepsilon_k/\partial k_x^2- \partial^2\varepsilon_k/\partial k_y^2)$ and $B_{2g}$ where $\gamma_{ab}=\partial^2\varepsilon_k/\partial k_x \partial k_y$.  The $B_{1g}$ geometry highlights the zone-face $(0,\pi)/(\pi,0)$ regions while the $B_{2g}$ geometry highlights the zone-diagonals $(\pm \pi/2,\pm\pi/2)$. 
This matrix element is appropriate for non-resonant Raman scattering and is the simplest one which is consistent with the symmetry. 
In practice incident laser frequencies are often chosen to take advantage of resonant enhancements arising from other degrees of freedom in the solid. These will change the absolute and relative magnitudes but not the symmetries of the vertices.

\item Quasiparticle contribution to  in-plane optical conductivity 
\begin{eqnarray}
\sigma(\Omega)&=&2\sum_K\int_K\frac{d^2k}{(2\pi)^2}\int\frac{d\omega}{\pi}\frac{f(\omega)-f(\omega+\Omega)}{\Omega}\times
\nonumber \\
&&v_{x}\text{Im}G_K(k,\omega+\Omega)v_{x} \text{Im}G_K(k,\omega)
\label{sigmadef}
\end{eqnarray}
where  $v_x=\partial \varepsilon_k/\partial k_x$. An approximate vertex correction \cite{Lin09} was also incorporated. 
\item Real and imaginary parts of sector-dependent electron self-energy $\Sigma_K(\omega)$.
\end{enumerate}
Here $\int_K$ denotes an integral over momenta lying in  sector $K$. Note that for the Raman scattering and in-plane optical conductivities (unlike for the other quantities we have considered) a vertex correction contribution (which we have only partially calculated) is present. The full vertex correction calculation  is currently in progress using  methods outlined in our previous work \cite{Lin09} but based on these results we expect the low frequency conductivity of primary interest here to have only a small vertex correction.

\section{Electron Spectral Function \label{Spectral}}

\begin{figure}[htbp]
\centering
\includegraphics[angle=-0,width=0.95\columnwidth]{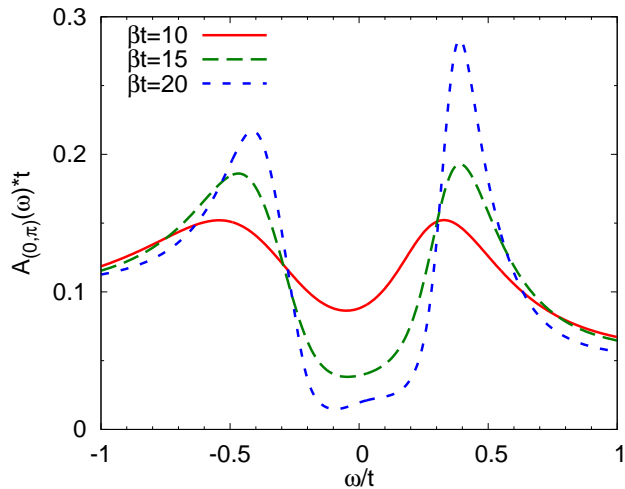}
\caption{Many-body density of states $A(\omega)=-\frac{1}{\pi}\text{Im}G(\omega)$ averaged over sector $C$ containing the $(0,\pi)$ point, calculated using 8-site DCA approximation at hole doping $x=0.05$ with $U=7t$ and inverse temperatures $\beta=1/T$ indicated.}
\label{spectral}
\end{figure}

Fig.~\ref{spectral} shows the electron spectral function for the sector containing the $(0,\pi)$ momentum (labeled $C$ in Fig.~\ref{tiling})  calculated  at $0.05$ hole doping for several temperatures. A `pseudogap' (reduction in density of states) is  visible in the low energy region. 

We define the pseudogap magnitude $E_{PG}=2\Delta_{PG}$ as the peak to peak separation (for the $x=0.05$ case shown in Fig.~\ref{spectral}  $E_{PG}\approx 0.8t\approx 0.3\ \text{eV})$.  The reduction in density of states is largest at the lowest temperature and for frequencies near $\omega=0$. It appears that at this doping the low frequency density of states vanishes as $T\rightarrow 0$.   As temperature is raised the gap `fills in': the density of states inside the gap increases but the gap magnitude does not change appreciably. For temperatures greater than about $0.2t$ the gap is no longer visible at this doping. 

\begin{figure}[htbp]
\centering
\includegraphics[angle=-0,width=0.95\columnwidth]{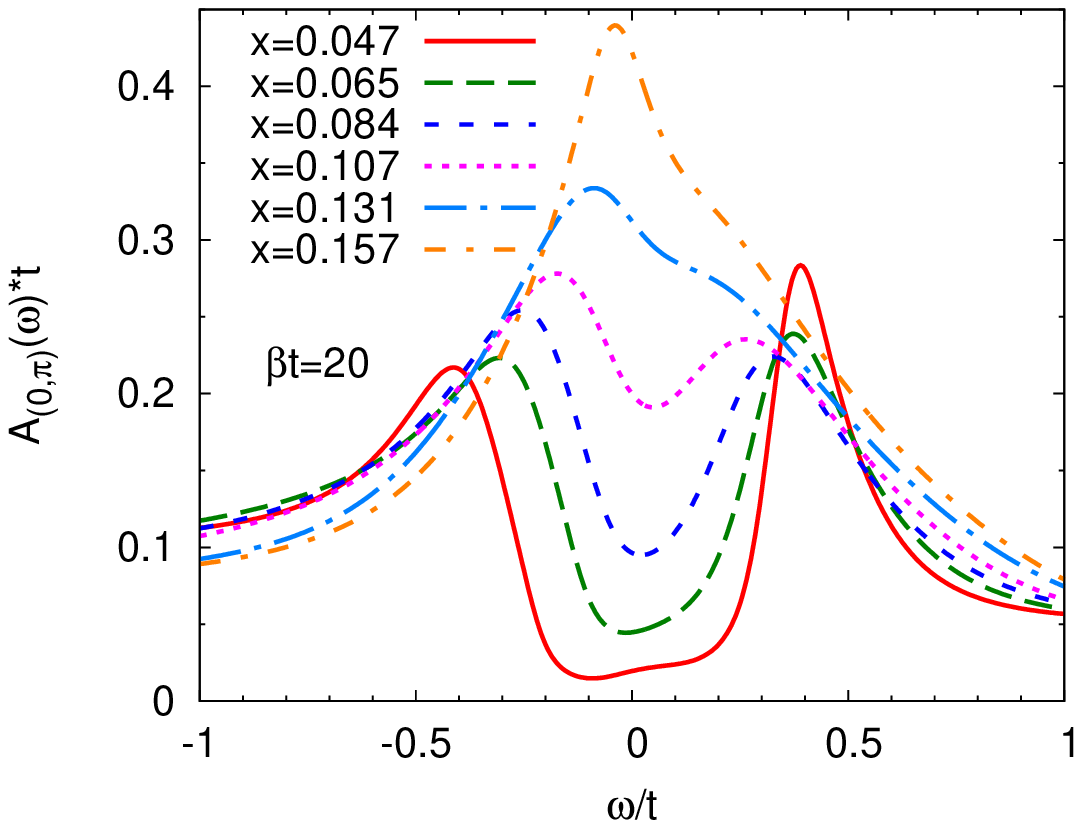}
\includegraphics[width=0.95\columnwidth]{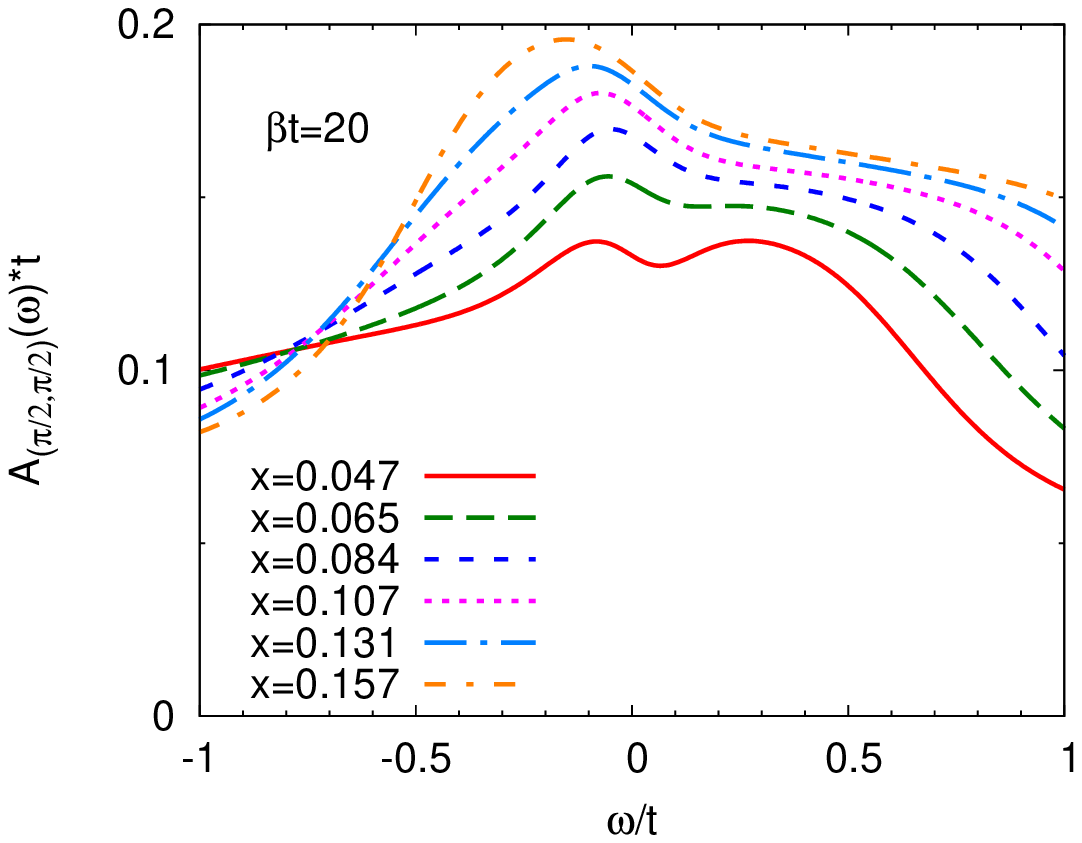}
\caption{Many-body density of states $A(\omega)=-\frac{1}{\pi}\text{Im}G(\omega)$ averaged over sector $C$ containing the $(0,\pi)$ point (upper panel) and sector $B$ containing the $(\pi/2,\pi/2)$ point (lower panel), calculated using the 8-site DCA approximation at hole dopings indicated with $U=7t$ and inverse temperature $\beta=20/t\approx 200\ \text{K}$.}
\label{spectral_doping}
\end{figure}

The upper panel of Fig.~\ref{spectral_doping} shows the $(0,\pi)$-sector spectral function calculated at several different dopings. A decrease in gap magnitude with increasing doping is evident. For dopings larger than $x=0.11$ a gap is not visible at the temperatures $T\geq t/20$ accessible to us although a weak feature in the $x=0.13$ curve suggests that the gap is still present.   However, certainly at $x=0.11$ and perhaps at $x=0.13$ the gap magnitude (as defined by the peak-to-peak distance in the spectral function) is not small. We therefore suspect that  at least a reduction of density of states would be observed at higher dopings if we were able to perform the calculations at lower temperatures. 

The lower panel of Fig.~\ref{spectral_doping} shows the $(\pi/2,\pi/2)$-sector spectral function at the same dopings. At the smallest doping a weak suppression of low frequency density of states is evident but for most dopings this sector remains ungapped.

\section{Interplane Conductivity \label{caxis}}

\begin{figure}[tbph]
\centering
\includegraphics[width=0.75\columnwidth,angle=-90]{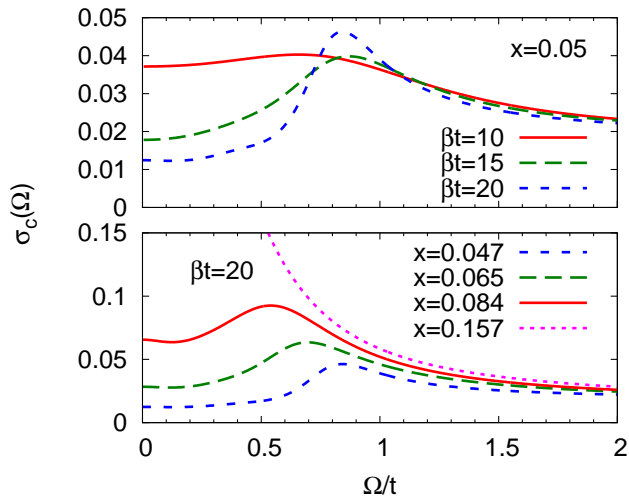}
\caption{Upper panel: temperature dependence of  interplane  conductivity $\sigma_c(\Omega)$ calculated from Eq.~(\ref{sigmacdef}) for hole doping $x=0.05$ at inverse temperatures $\beta=1/T$ indicated. Lower panel (note change of vertical scale): doping dependence of interplane conductivity calculated at inverse temperature $\beta=20/t\approx 200\ \text{K}$. At hole doping $x=0.16$, the low frequency interplane conductivity is found to be approximately of Lorentzian form with dc value $0.65$.}
\label{caxistempanddoping}
\end{figure}

An important early indication of the presence of a charge-pseudogap was provided by measurements of the frequency dependence of the  interplane conductivity.\cite{Homes93} As can be seen from Eq.~(\ref{sigmacdef}), in high-$T_c$ materials the matrix elements relevant to the interplane conductivity highlight the zone-face regions where the electron spectral function exhibits a gap (see upper panel in Fig.~\ref{spectral_doping}). 

Fig.~\ref{caxistempanddoping} shows the calculated temperature and doping dependence of the  interplane conductivity. The pseudogap is visible as a temperature and doping dependent suppression of the low frequency interplane conductivity.  The interplane conductivity is suppressed over a relatively wide frequency range; the suppression increases as the doping or temperature decreases, and the gap fills in but does not close as temperature is increased.  The calculations also reveal a weak maximum in the conductivity at an energy just above the pseudogap. A somewhat broader version of this feature was observed by Yu {\it et al.}\cite{Yu08} It is possible that the relative sharpness of the feature  is an artifact related to our coarse-graining of momentum space, which might arise because the DCA approximation necessarily produces a gap that is piecewise continuous; and as is known from the familiar case of s-wave BCS superconductivity a momentum independent gap produces a peak. The results are reasonably consistent with experiment.\cite{Homes93,Tajima97,Basov06,Yu08}  Ref.~\onlinecite{Yu08} reports a high energy pseudogap of a magnitude consistent with what is found here.
  It is important to note that in the widely studied $\text{YBa}_2\text{Cu}_3\text{O}_{6+x}$ material the interplay of strong local field effects (arising from the bilayer structure) and phonon effects produce complicated structures in the low frequency conductivity which are not represented in the present calculation.\cite{Dulic01,Shah01,Yu08}  

Conductivities may be characterized by `spectral weight', the integrated area in some frequency range. The total spectral weight obeys an `f-sum' rule, which  for the model studied here is
\begin{equation}
\begin{aligned}
\int_0^\infty \frac{2d\omega}{\pi}\sigma_c(\omega)=2\sum_K\int_K&\frac{d^2k}{(2\pi)^2}\int_{-\infty}^{\infty}\frac{d\omega}{\pi}\\
&f(\omega)t_\perp^2(k)\text{Im}[G^2(k,\omega)].
\end{aligned}
\label{caxissumrule}
\end{equation}
We have verified that the spectral weight obtained from integration of the conductivity equals the spectral weight obtained from an evaluation of Eq.~(\ref{caxissumrule}). The curves shown in Fig.~\ref{caxistempanddoping} imply a temperature and doping-dependent decrease in low frequency spectral weight. We find that in the lower doping regions where the pseudogap is present,  the spectral weight which is lost at low frequencies due to the formation of the gap  is not fully restored at higher frequencies;  thus the conductivity spectral weight decreases as the temperature or doping is decreased, whereas at higher doping the c-axis conductivity spectral weight increases as temperature is decreased. The calculated interplane conductivity is spread over a wide frequency range so the pseudogap-induced decrease of spectral weight is small relative to the total weight.  The doping and temperature dependence of the c-axis spectral weight is given in Table \ref{sumruletable}.

Ref.~\onlinecite{Ferrero10} reports interplane conductivity  results obtained using a two-site cluster, a particular choice of self-energy periodization and a Pad\'{e} continuation to compute $\sigma_c$ for $U/t=10$ and $t'/t=-0.3$. The results are very similar to those shown here but with slightly larger gaps and a much greater suppression of $\sigma_c$ at sub gap frequencies.  Note that  a factor of $\pi t_0^2/(8t^2)$ is required to convert the  results of Ref.~\onlinecite{Ferrero10} to our units.

\begin{table}[tbph]
\begin{tabular}{cccc}
\hline\hline
x& $\beta t=10$ & $\beta t=15$ & $\beta t=20$\\
\hline
0.05 & 0.111 & 0.105 & 0.102\\
0.07 & 0.125 & 0.119 & 0.116\\
0.08 & 0.142 & 0.138 & 0.134\\
0.16 & 0.207 & 0.223 & 0.231\\
\hline\hline
\end{tabular}
\caption{Integrated spectral weight of  interplane optical conductivity obtained from Eq.~(\ref{caxissumrule}) at different temperatures for several hole dopings.}
\label{sumruletable}
\end{table}

To conclude this section we note that the pseudogap effects are enhanced by the structure of the matrix element (which is the one widely used in 
the theoretical literature) which enhances the contribution of
sector $C$ relative to sector $B$. If a $k$-independent matrix element was used, the pseudogap effects would be much less pronounced.

\section{Raman Scattering Intensity \label{Raman}}

\begin{figure}[htbp]
\centering
\includegraphics[width=0.95\columnwidth,angle=-0]{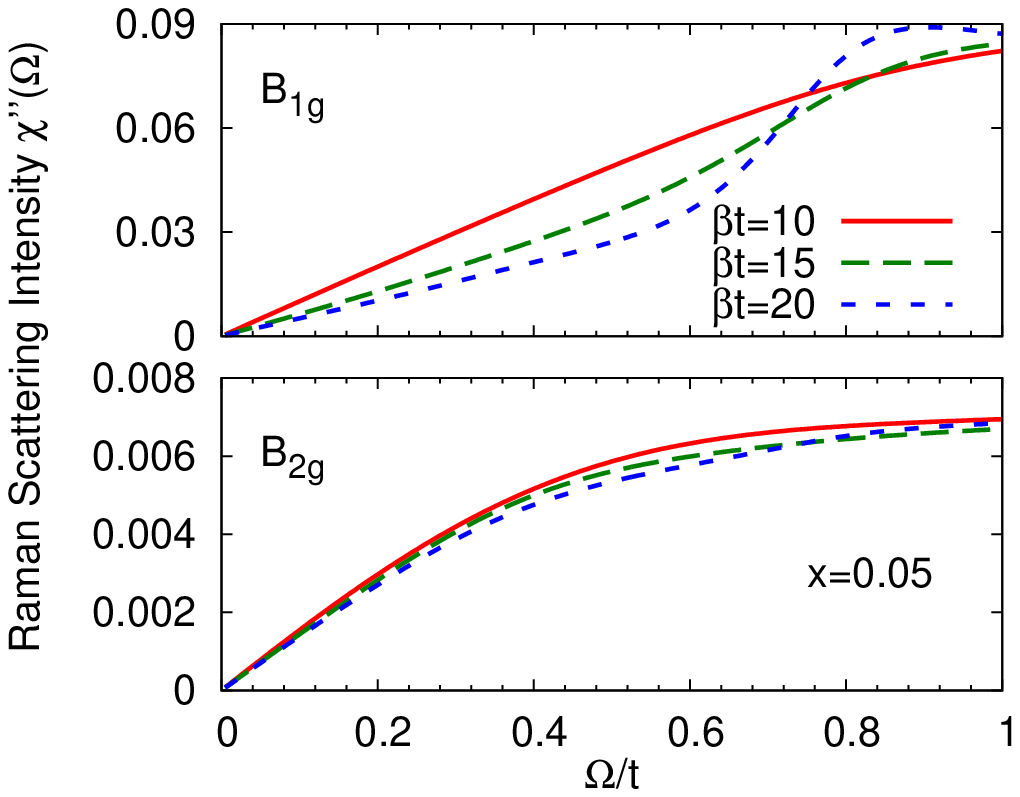}
\includegraphics[width=0.95\columnwidth,angle=-0]{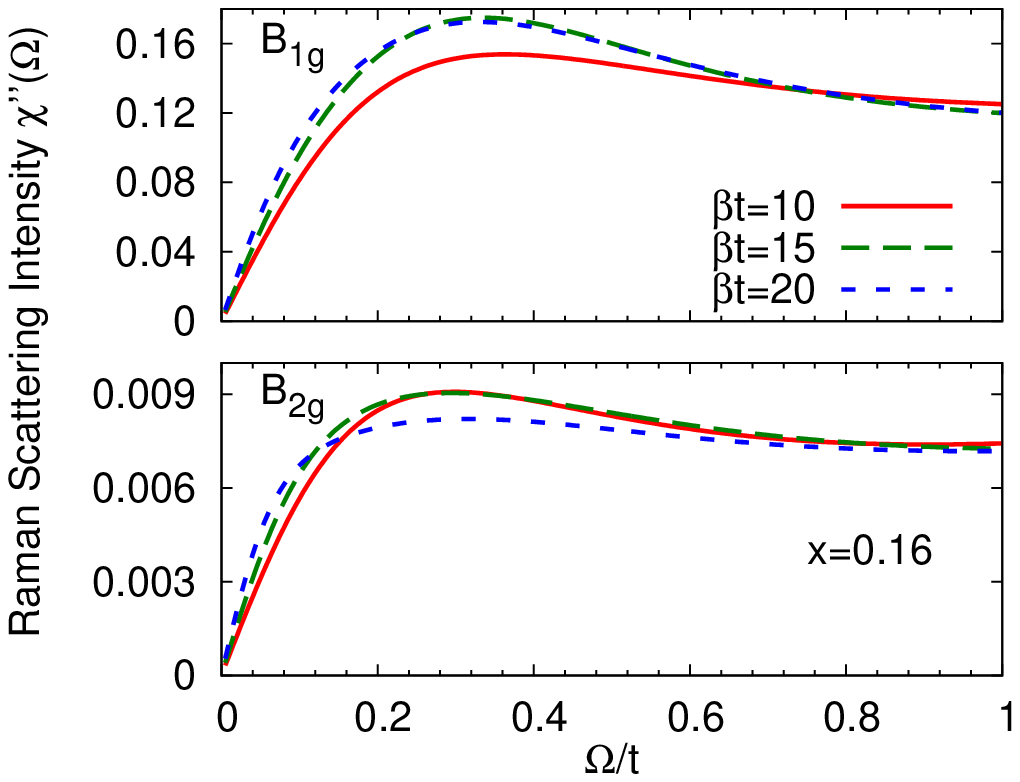}
\caption{Temperature dependence of Raman scattering intensity $\chi''(\Omega)$ calculated from Eq.~(\ref{ramandef}) for $B_{1g}$ and ${B_{2g}}$ scattering channels in 8-site DCA at hole doping  0.05 (upper panels) and 0.16 (lower panels) and $U=7t$.}
\label{raman05}
\end{figure}
Raman scattering has provided important information about the cuprates.\cite{Devereaux07} It is a photon-in/photon-out process, and the polarizations of the electric field vectors of the incident and emitted photons may be adjusted to highlight transitions in different regions of momentum space. We consider here two scattering geometries: $B_{1g}$ which highlights the anti-nodal region, and $B_{2g}$ where the matrix element highlights the nodal region.  The computation of the Raman intensity involves a vertex correction, which we have not computed (note that unlike the case of the in-plane conductivity \cite{Lin09} the vertex correction does not involve a contribution from the momentum-space discontinuity of the self-energy). Thus what is shown is the quasiparticle contribution only. 
For this reason, and also because we do not have control over the high frequency part of the spectrum, we do not undertake the sum rule analysis introduced by de Medici {\it et al.}\cite{deMedici08} in their single site DMFT study of the Raman intensity in the strong correlation limit.


Fig.~\ref{raman05} shows the temperature dependence of the calculated Raman spectra in the two scattering geometries at the low doping $x=0.05$ (upper panels) and high doping $x=0.16$ (lower panels). We see that at the higher dopings the functional forms and temperature dependencies in the two scattering channels are similar, while at the lower doping they are rather different. At the higher doping the Raman response in each channel rises linearly to a weak maximum and then approximately saturates; the initial slope increases as $T$ decreases in both channels. It is also interesting to note that the calculation reproduces  the roughly frequency-independent high frequency behavior, which had previously been argued to be evidence for novel `marginal Fermi liquid' physics not contained in the Hubbard model.\cite{Varma89} However, at the lower dopings the low frequency $B_{1g}$  response is progressively suppressed as $T$ decreases, whereas the $B_{2g}$ response has negligible $T$ dependence.

 The key features of the results presented in Fig.~\ref{raman05}, namely an increase  with increasing doping in the temperature dependence of the  low frequency $B_{2g}$ scattering intensity and a change in sign of the $T$ dependence for the $B_{1g}$ intensity along with the presence of a weak maximum at an energy which decreases with increasing doping are qualitatively consistent with data (see e.g. Fig.~42 and 43 of Ref.~\onlinecite{Devereaux07}).

However, there are important quantitative differences.
The calculated ratio of $B_{1g}$ to $B_{2g}$ is approximately $15$, whereas the experimental ratio is much closer to one. This difference presumably arises because the experiments employ a resonantly enhanced matrix element.
More importantly, the calculation places the maximum (or scale at which the response saturates) at a higher energy than the data in Refs.~\onlinecite{Blanc09} and \onlinecite{Devereaux07}. For example at $x=0.16$ the calculated saturation point for the ${B_{1g}}$ spectra (using $t=0.35\ \text{eV}$ and placing the saturation maxima at $0.33t$) at $\beta t=20$ is $\omega \approx 925\ \text{cm}^{-1}$ while in the ${B_{2g}}$ spectra our results saturate at $\approx 700 \ \text{cm}^{-1}$ (placing the saturation point at $0.25t$) whereas the experimental data have not quite saturated. Further, observed ${B_{2g}}$ Raman spectra at low dopings have more $T$ dependence than is found in our calculations. The onset coincides in temperature with other measures of the pseudogap, so we associate the phenomenon to the pseudogap.

\begin{figure}[htbp]
\centering
\includegraphics[width=0.95\columnwidth]{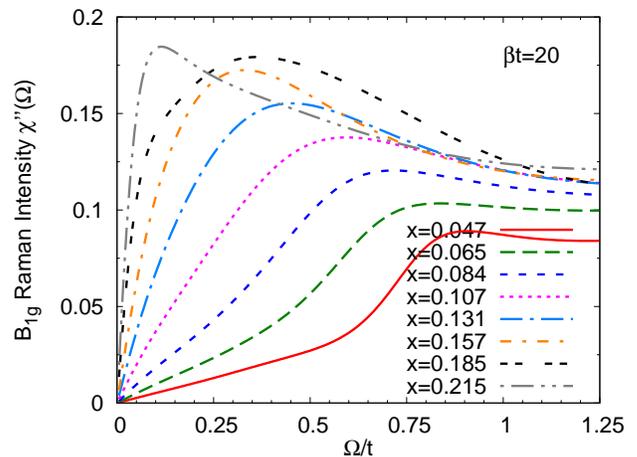}
\caption{Raman scattering intensity $\chi''(\Omega)$ in the $B_{1g}$ geometry at different hole dopings as indicated with $U=7t$ and inverse temperature $\beta=20/t\approx200\ \text{K}$.}
\label{B1gdoping} 
\end{figure}

\begin{figure}[htbp]
\centering
\includegraphics[width=0.95\columnwidth]{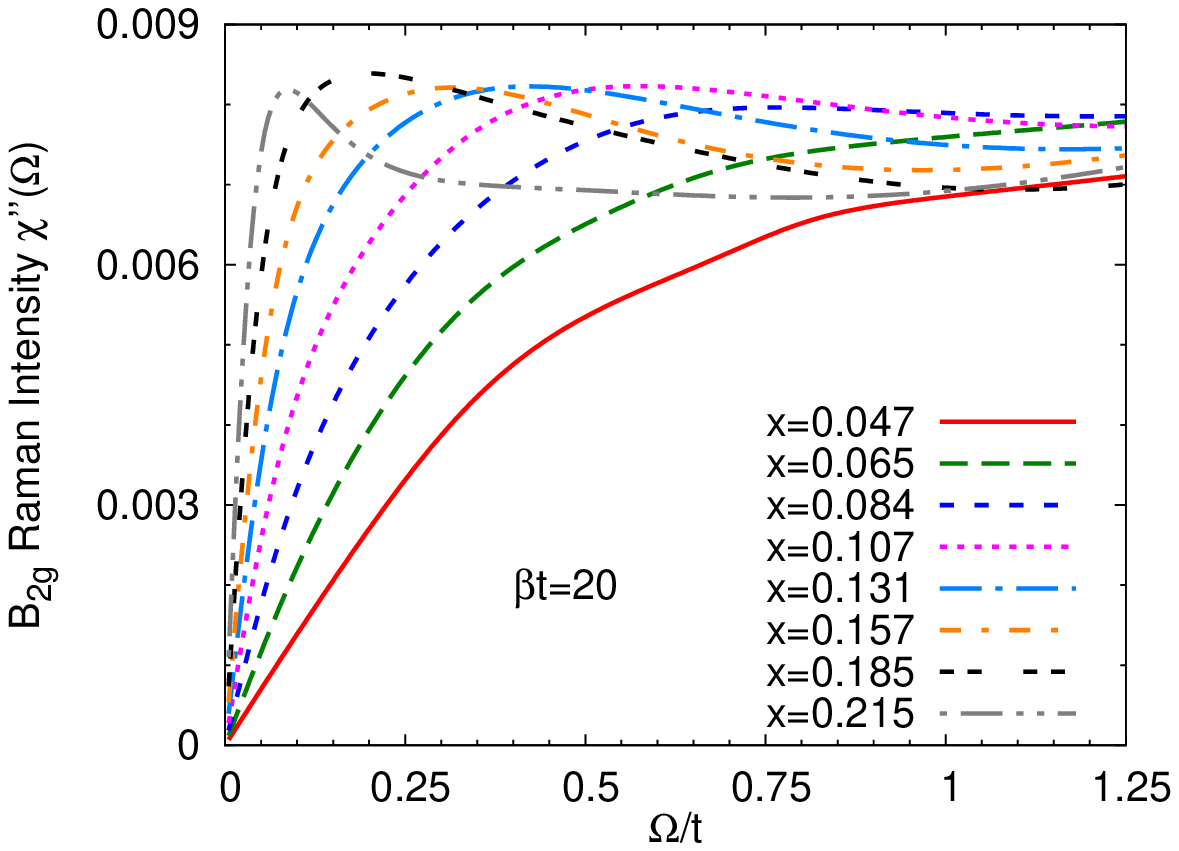}
\caption{Raman scattering intensity $\chi''(\Omega)$ in the $B_{2g}$ geometry at different hole dopings as indicated with $U=7t$ and inverse temperature $\beta=20/t\approx200\ \text{K}$.}
\label{B2gdoping}
\end{figure}

Figs.~\ref{B1gdoping} and \ref{B2gdoping} show the doping dependence of the Raman intensity at inverse temperature $\beta t=20$ corresponding to $T \approx 200\ \text{K}$.
A strong increase in initial slope is evident in both channels.
The rapid steepening of the initial slope in the $B_{2g}$ channel is a consequence of the emergence of coherent quasiparticles in the zone-diagonal sector (see Sec.~\ref{Selfenergy}).
The change in the $B_{1g}$ channel arises from the doping dependence of the pseudogap.
The calculated behavior of the Raman intensity, including the difference in doping and temperature dependence  between the two sectors and in particular the doping dependent suppression of the $B_{1g}$ Raman spectra over a wide frequency range, is in reasonable agreement with measurements.\cite{Katsufuji94,Venturini02,Hackl05,Devereaux07}
However, our calculated $B_{2g}$ spectra exhibit a strong doping dependence which is not observed in recent experiments.\cite{Blanc09}
%
%

\section{In-plane conductivity\label{conductivity}}

\begin{figure}[htbp]
\centering
\includegraphics[angle=-0,width=0.95\columnwidth]{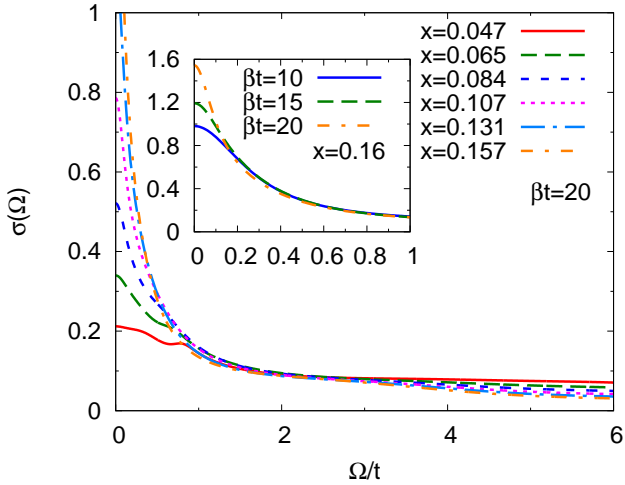}
\caption{Main panel: in-plane optical conductivity $\sigma(\Omega)$ calculated from Eq.~(\ref{sigmadef}) at $U=7t$ and inverse temperature $\beta=20/t\approx 200\ \text{K}$ for hole dopings indicated. Inset: temperature dependence of the low frequency in-plane optical conductivity for hole doping $x=0.16$.}
\label{Conductivitydoping}
\end{figure}

The optical conductivity $\sigma(\Omega)$ of  the high-$T_c$ cuprates has been an enduring mystery.  The salient features of the data are a `Drude' peak centered at zero frequency, with a strongly temperature dependent spectral weight and half width, and a broad higher frequency continuum.\cite{Orenstein91,Rotter91,Basov06} The strong doping dependence of the `Drude' peak has been taken as evidence of strong `Mott' correlations, while the broad higher frequency continuum has been interpreted in terms of scattering from spin fluctuations.\cite{Varma89,Jaklic00,Abanov01b,Schachinger03}

\begin{figure}[htbp]
\centering
\includegraphics[angle=-0,width=0.95\columnwidth]{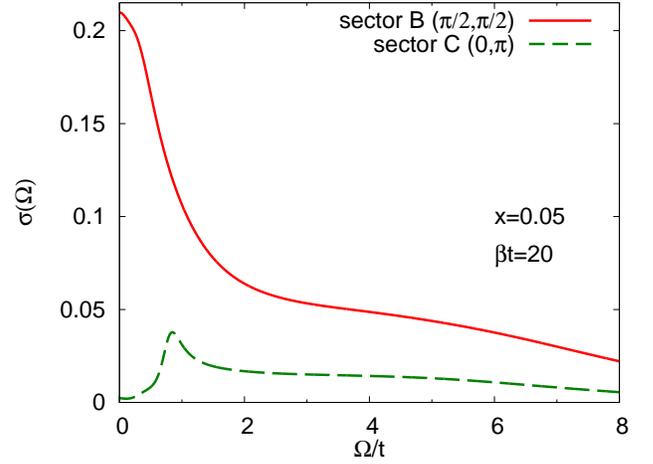}
\caption{Contribution of different momentum sectors to calculated in-plane optical conductivity for $U=7t$ at doping $x=0.05$ and inverse temperature $\beta =20/t \approx 200\ \text{K}$.}
\label{Conductivitysector}
\end{figure}

The computation of the in-plane conductivity involves a vertex correction \cite{Lin09} which has not been fully calculated. Here we only include the vertex correction arising from the self-energy discontinuities in momentum-space, while the contribution of additional dependence of  the self-energy on the vector potential $\vec{A}$ which arises from the change in mean-field function is neglected. 

The main panel of Fig.~\ref{Conductivitydoping} presents  the calculated doping dependence of the real part of the in-plane conductivity. Our results exhibit strong similarities to experimental data, including, at low frequencies, a growth of the `Drude' peak with doping, and a weakly frequency and doping-dependent `mid-infrared' conductivity with a magnitude comparable to the measured value $\sim 5-700\ \Omega^{-1}\text{cm}^{-1}$ (note that to convert to the unit of $\Omega^{-1}\text{cm}^{-1}$ commonly used in experiments, our results have to be multiplied by a factor of about $5\times10^3$). The inset shows that at higher dopings the `Drude' peak grows noticeably and sharpens slightly as temperature is decreased. 

\begin{table}[tbph]
\begin{tabular}{cccc}
\hline\hline
x& $\beta t=10$ & $\beta t=15$ & $\beta t=20$\\
\hline
0.05 & 0.252&  0.244 & 0.240\\
0.08 & 0.3141& 0.3145 &   0.313\\
0.11 &   0.328& 0.343 &   0.345\\
0.16 & 0.383&   0.390 &  0.392\\
\hline\hline
\end{tabular}
\caption{Integrated spectral weight of  in-plane optical conductivity obtained by integrating calculated curves multiplied by a factor of $2/\pi$ up to $\Omega=3t\approx 1\ \text{eV}$  at different temperatures for several hole dopings.}
\label{inplanesumruletable}
\end{table}

A particularly interesting issue is the lack of a clearcut effect of the pseudogap on the conductivity. A natural conjecture is that the conductivity is dominated by states near the zone-diagonal, which are not sensitive to the pseudogap.  Fig.~\ref{Conductivitysector} presents a decomposition of the contributions of the different sectors to the measured conductivity which supports this conjecture. A pseudogap (the feature at $\omega\sim t$)  is present in the contribution of sector $C$, but the contribution of this sector to the measured conductivity is relatively small, so the feature is not evident in the full conductivity. The pseudogap feature is more pronounced in $4$-site cluster calculations \cite{Haule07,Chakraborty08,Lin09} because the geometry of the 4-site cluster is such that the low frequency behavior is dominated by the $(0,\pi)/(\pi,0)$ sectors so that it does not capture the physics of the nodal quasi-particles. In the present 8-site calculation the nodal quasiparticle physics is represented by sector $B$, which is seen to give the dominant contribution to the conductivity.

The doping \cite{Uchida91,Orenstein91,Rotter91,Comanac08}  and temperature \cite{Santander03,Carbone06} dependence of the optical spectral weight $K=\int \frac{2d\omega}{\pi} \sigma(\omega)$ has been the subject of discussion in the literature.  One ambiguity is the range over which the conductivity is to be integrated:  setting the range too high leads to the inclusion of interband transitions which are believed to be irrelevant to the physics of high-$T_c$ materials while setting the range too low may mean that changes in the width of a low frequency peak may be mistaken for changes in area.  Recent papers suggest that an upper cutoff of $\sim 1\ \text{eV}$ is a  reasonable compromise value.\cite{Millis05,Carbone06} A marked increase of spectral weight  occurs as doping is increased (see Ref.~\onlinecite{Comanac08} for a summary of the data). The experimental consensus is that at all dopings the spectral weight increases weakly as temperature is decreased, and that the temperature dependence changes markedly as the temperature is decreased below  the superconducting transition temperature. The doping dependence of the calculated spectral weight over the range $\Omega<3t\sim 1\ \text{eV}$ (shown in Table \ref{inplanesumruletable}) is reasonably consistent with the data although we find that at lower dopings $x=0.08$, $0.05$ the temperature dependence flattens.
A weak temperature dependence is seen in the spectral weights integrated up to $3t \approx 1\ \text{eV}$. The sign of the temperature dependence changes with doping: increasing as $T$ is decreased at high doping and decreasing as $T$ decreases at low doping.
It is tempting to relate this change in temperature dependence to a change in physics from Fermi-liquid-like at high doping to pseudogapped at low doping
but the small magnitude of the effect and the possibility of temperature dependent changes in the functional form of the conductivity make
an interpretation unclear.
Calculations of a $4$-site cluster approximation to the $t-J$ model\cite{Carbone06} (which is believed to reflect the low energy physics of the Hubbard model 
for $U\geq 12t$, larger than what we studied here) found a similar doping and temperature dependence.

 \begin{figure}[htbp]
\centering
\includegraphics[angle=-0,width=0.95\columnwidth]{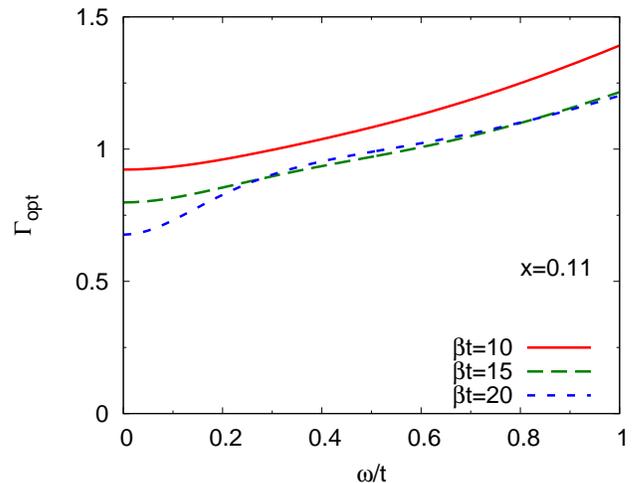}
\caption{Optical scattering rate $\Gamma_{\text{opt}}$ computed from Eq.~(\ref{gamdef})  for hole doping $x=0.11$ and inverse temperatures $\beta=1/T$  indicated.}
\label{gammaopt}
\end{figure}

The conductivity is sometimes expressed in terms of a frequency-dependent optical scattering rate $\Gamma_{\text{opt}}$ 
related to the complex conductivity ${\tilde \sigma}= \sigma_1(\omega)+i\sigma_2(\omega)$ via 
 \begin{equation}
 \begin{aligned}
 \Gamma_{\text{opt}}&=K\text{Re}\frac{1}{\tilde{\sigma}}
 \end{aligned}
\label{gamdef}
\end{equation}
where $\sigma_1$ is calculated from Eq.~(\ref{sigmadef}) and $\sigma_2$ is obtained by the Kramers-Kronig transformation of $\sigma_1$.
In underdoped cuprates at high temperature, $\Gamma_{\text{opt}}$ is large and temperature dependent over a wide frequency range. As the temperature is decreased to the pseudogap scale the high frequency part loses its temperature dependence while at lower frequencies a temperature dependent suppression of $\Gamma_{\text{opt}}$ appears.\cite{Timusk95} Fig.~\ref{gammaopt} shows that this behavior is also found in our calculations.

\section{Electron self-energy\label{Selfenergy}}

\begin{figure}[htbp]
\centering
\includegraphics[width=0.95\columnwidth,angle=-0]{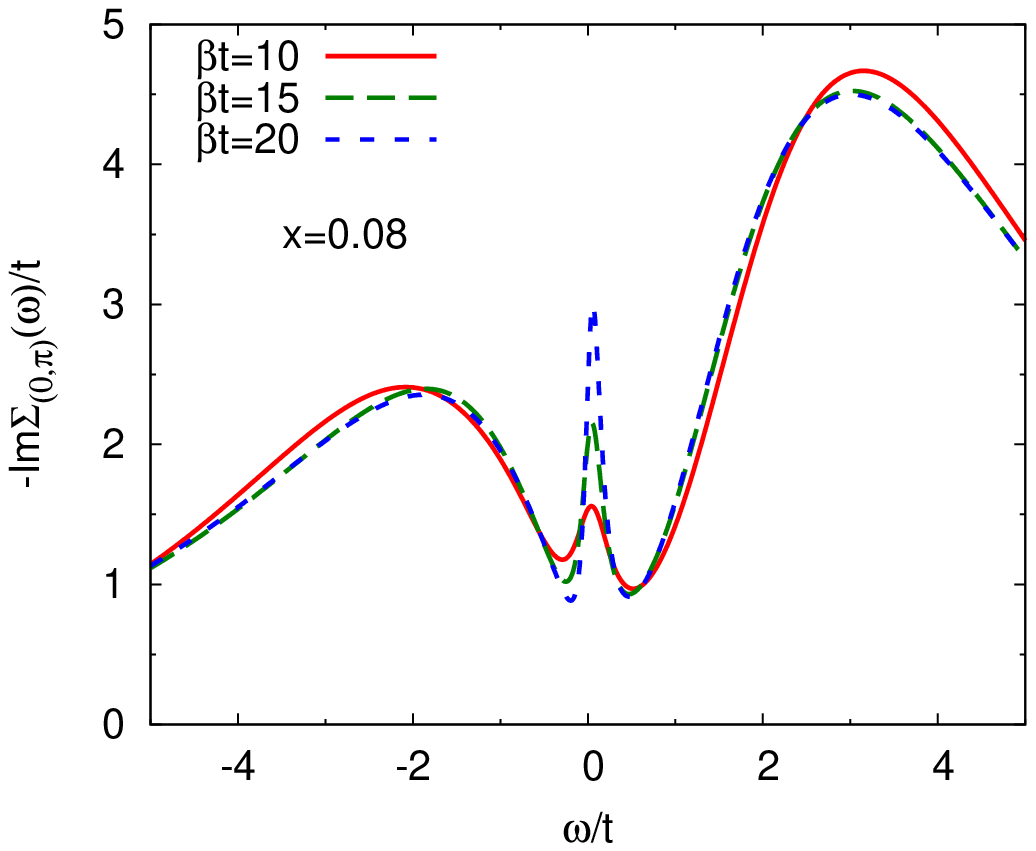}
\includegraphics[width=0.95\columnwidth,angle=-0]{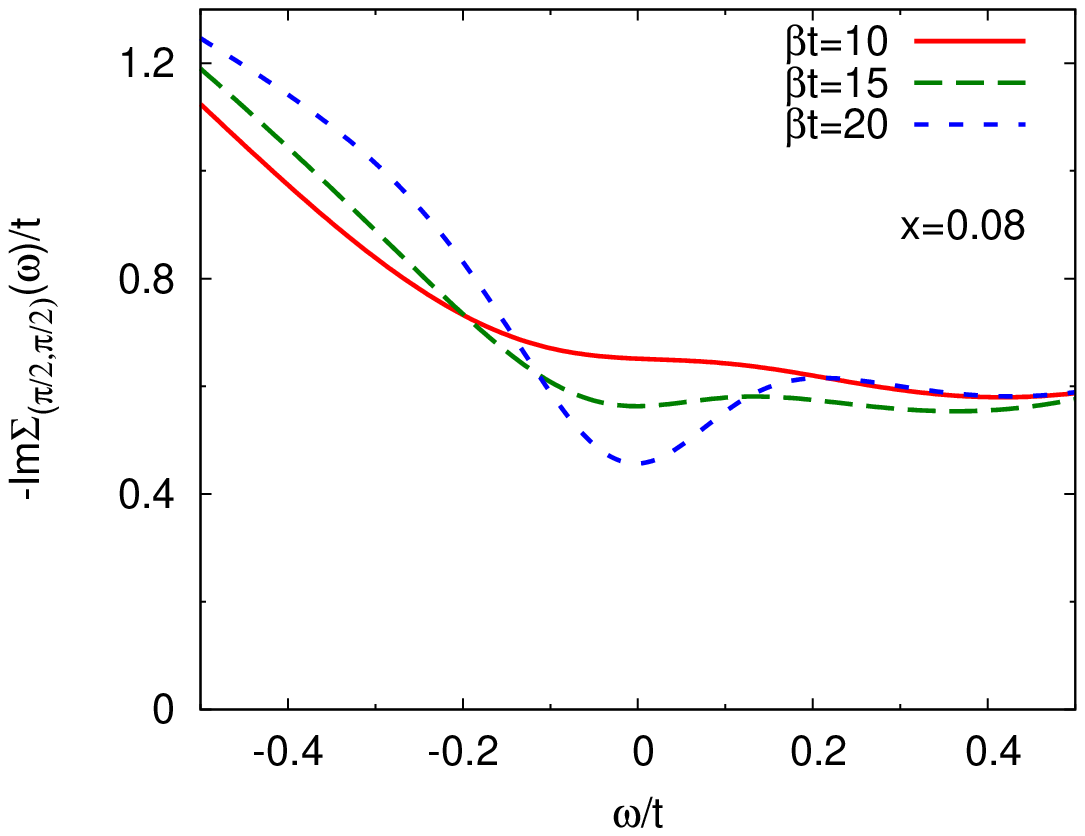}
\caption{Imaginary part of electron self-energy computed at different temperatures for hole doping $x=0.08$ in $(0,\pi)$-sector (upper panel) and $(\pi/2,\pi/2)$-sector (lower panel).}
\label{selfenergy_temp}
\end{figure}
\begin{figure}[htbp]
\centering
\includegraphics[width=0.95\columnwidth,angle=-0]{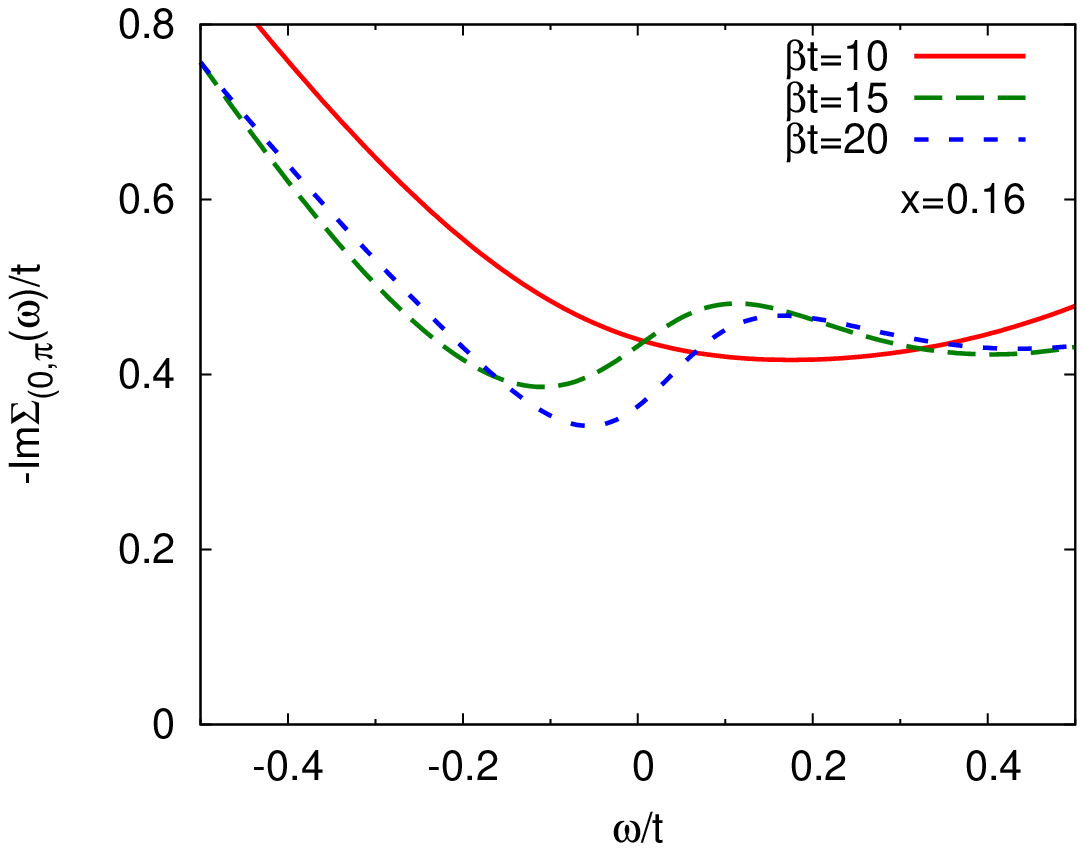}
\includegraphics[width=0.95\columnwidth,angle=-0]{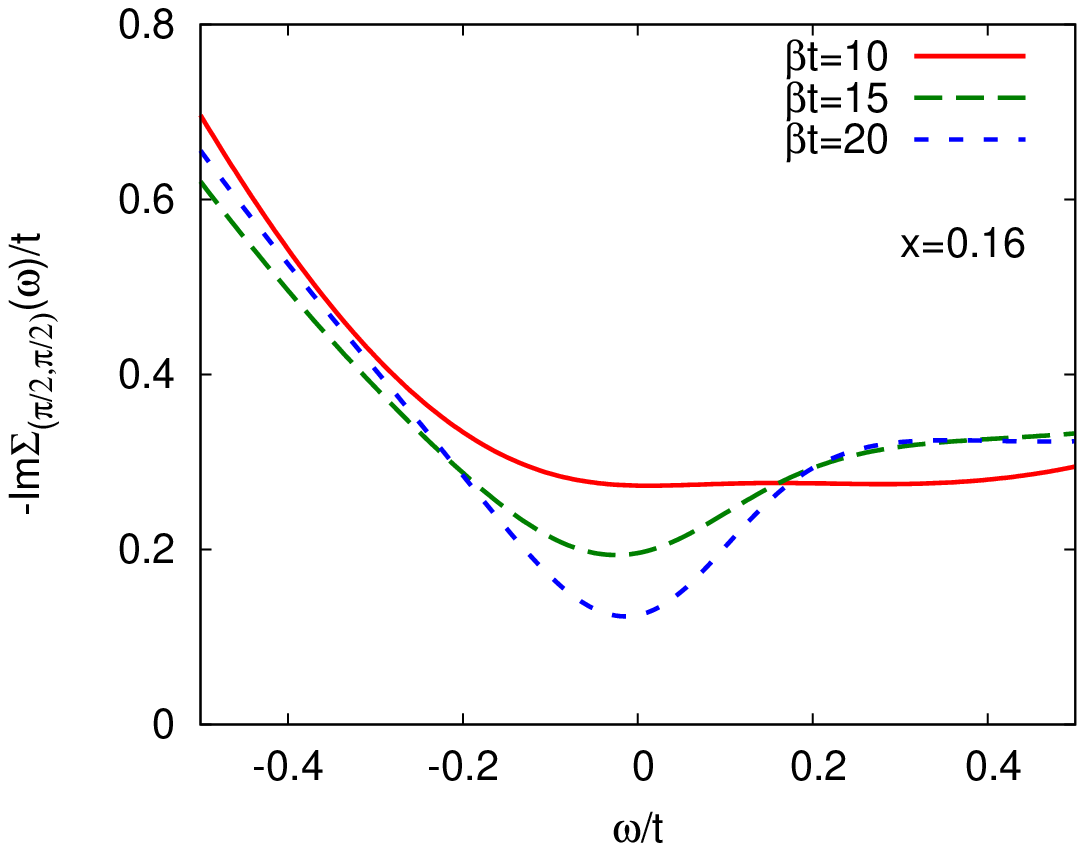}
\caption{Imaginary part of electron self-energy computed at different temperatures for hole doping around $0.16$ in $(0,\pi)$-sector (upper panel) and $(\pi/2,\pi/2)$-sector (lower panel).}
\label{selfenergy_temp_x0.16}
\end{figure}

Fig.~\ref{selfenergy_temp} and~\ref{selfenergy_temp_x0.16} compare the temperature dependence of the imaginary part of the electron self-energy computed at doping $x=0.08$ and $x=0.16$ for the two momentum sectors containing the Fermi surface. At doping $x=0.08$ (see Fig.~\ref{selfenergy_temp}), we see that the $(0,\pi)$ sector self-energy is characterized by a pole located near $\omega=0$ whereas no pole appears in the self-energy of the sector containing the zone-diagonal part. A near zero-energy pole in the self-energy is a characteristic of a Mott insulating state, confirming that the gap opening transition is indeed a sector selective Mott transition. As the temperature is decreased the pole grows in strength. The pole position  at $\omega\approx 0$ leads to  the approximate particle-hole symmetry of the spectra. We also observe that at this doping the sector $B$ (zone-diagonal) self-energy has only a weak  temperature dependence at the temperatures accessible to us. 

At doping $x=0.16$ (see Fig.~\ref{selfenergy_temp_x0.16}), the self-energies in both sectors decrease with temperature and have a minimum centered at $\omega\approx 0$. In sector $B$, the self-energy has Fermi-liquid-like behavior which decreases rapidly and roughly linear in $T$. In sector $C$, the self-energy decreases more slowly.
The difference in magnitude and doping dependence indicates that this doping regime is characterized by a large variation in scattering rate around the Fermi surface.

\begin{figure}[htbp]
\centering
\includegraphics[angle=-0,width=0.95\columnwidth]{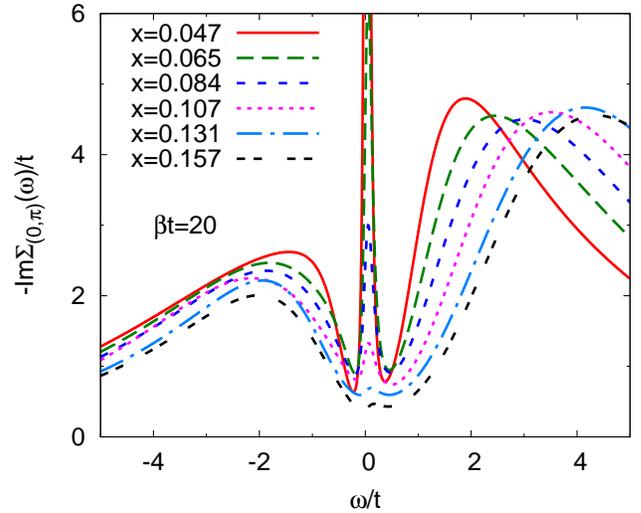}
\caption{Imaginary part of electron self-energy computed for sector $C$ containing the $(0,\pi)$ point at inverse temperature $\beta=20/t\approx200\ \text{K}$ for dopings indicated. (For the real part see Fig.~\ref{resigma}.)}
\label{ImSigC}
\end{figure}

Fig.~\ref{ImSigC} shows the doping dependence of the imaginary part of the  sector $C$ self-energy. A pole, gaining in strength as doping is decreased and centered at approximately zero frequency, is clearly visible for $x\leq 0.11$, while for $x=0.13$ and $0.16$ there is no pole, only a weak modulation indicating a non-Fermi-liquid scattering rate. Whether this modulation would evolve into a pole as $T\rightarrow0$ is an interesting open question.
Liebsch {\it et al.} \cite{Liebsch09} analysed the self-energy pole structure in the $(0,\pi)$ sector of a $4$-site CDMFT study, finding a similar doping dependence of the pole strength. They reported a strong dependence of the pole position on doping; this variation is not found in the $8$-site cluster studied here.

An alternative characterization of electronic behavior is the quasiparticle residue $Z = \left[1 - \left.\frac{\partial \text{Re}\Sigma(\omega)}{\partial \omega}\right|_{\omega=0}\right]^{-1}$.
Because within each sector the self-energy is momentum-independent,  $Z$ gives the renormalization of the Fermi velocity as $v^*= Z v$. 
This renormalization has physical significance if the self-energy is Fermi-liquid-like, meaning that the imaginary part is not too large and the real part is linear in frequency over a reasonable range about $\omega=0$. We determine the boundaries of the Fermi-liquid regime by first observing that the real part of the self-energy $\text{Re}\Sigma$ is linear in frequency over the range $-\omega_L\le\omega\le\omega_H$, and then comparing the magnitude of the imaginary part of the self-energy at zero frequency to the change of $\omega-\left(\text{Re}\Sigma(\omega)-\text{Re}\Sigma(0)\right)$ over the linear range. If the change
$(\omega_H-\text{Re}\Sigma(\omega_H))-(\omega_L-\text{Re}\Sigma(\omega_L))$ is larger than $-2\text{Im} \Sigma(\omega=0)$ we identify the regime as Fermi-liquid-like. 
For an illustration of the determination of Fermi-liquid behavior see appendix \ref{detflregime}.

This condition is   reasonably well satisfied for sector $B$ for dopings $x>0.08$ (and marginally satisfied for $x=0.08$). Similarly sector $C$ is found to be Fermi-liquid-like for dopings $x=0.18$ and greater, but for $x=0.06$ the self-energy in both sectors is far from Fermi-liquid-like and the quantity $Z$ cannot be interpreted as a ``quasiparticle weight''. 

\begin{figure}[htbp]
\begin{center}
\includegraphics[angle=-0,width=0.95\columnwidth]{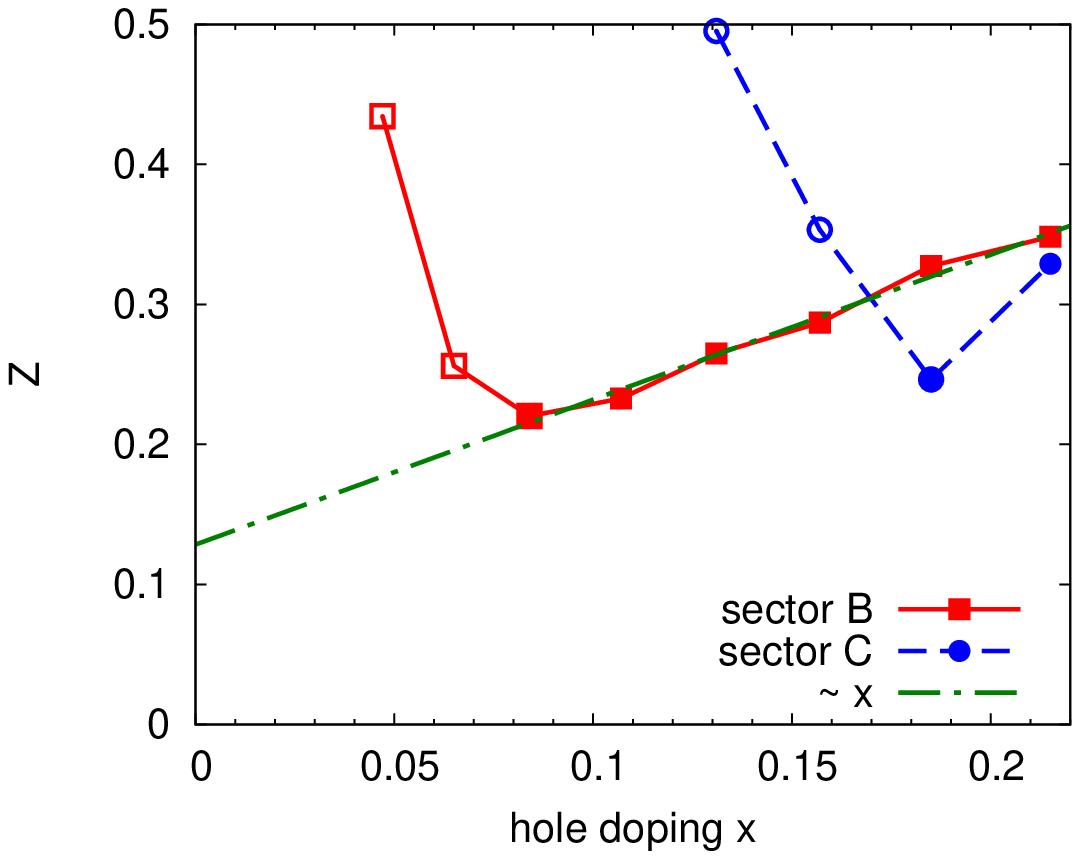}
\includegraphics[width=0.95\columnwidth]{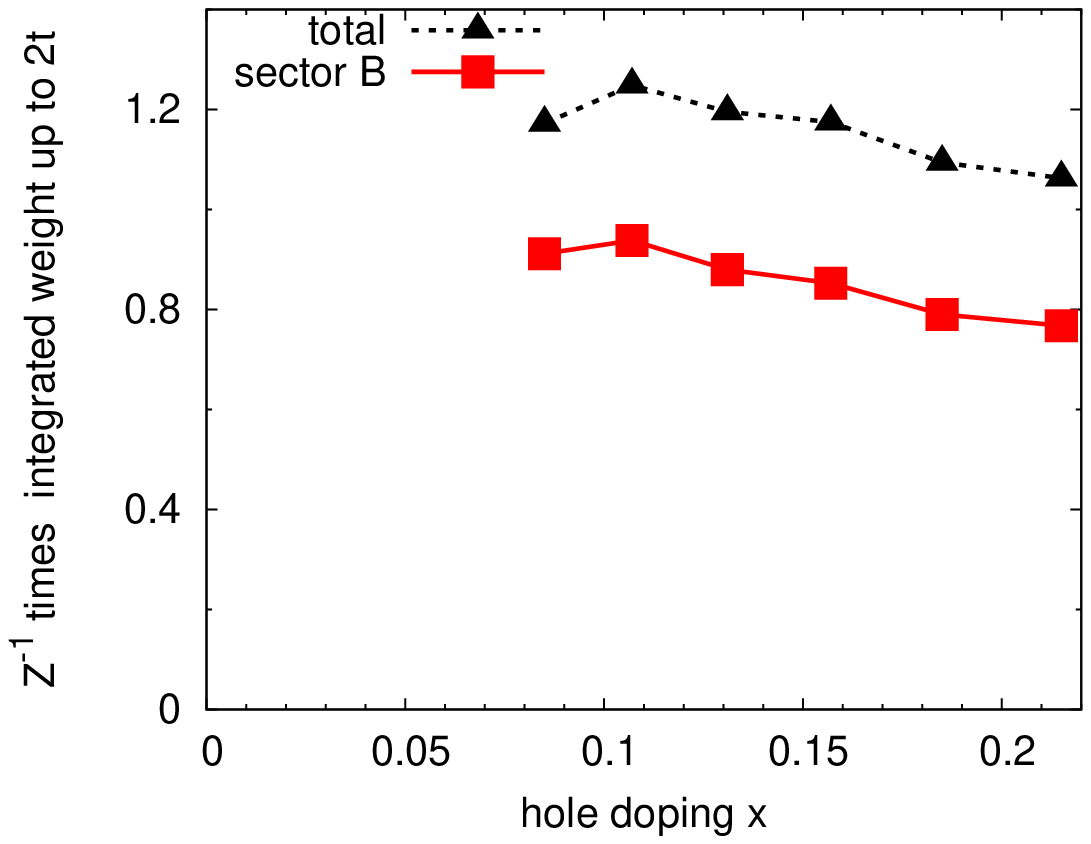}
\end{center}
\caption{Upper panel: doping dependence of the quasiparticle residue $Z$ calculated for sector $B$ containing the $(\pi/2,\pi/2)$ point and sector $C$ containing the $(0,\pi)$ point from analytically continued self-energies at inverse temperature $\beta=20/t\approx 200\ \text{K}$. The filled symbols represent dopings $x\geq0.08$ for sector $B$ and $x\geq0.18$ for sector $C$ where the self-energy is Fermi-liquid-like as defined in the text. Open symbols are mathematically defined from our data but we believe are not physically meaningful because the scattering rate is too large. Lower panel: approximate proportionality of optical spectral weight and $Z$. Trinagles (black online): integral from $0$ to $\omega=2t$ of calculated optical conductivities, divided by renormalization factor $Z$ of sector $B$, plotted against doping $x$. Squares (red online): integral from $0$ to $\omega=2t$ of sector $B$ contribution to conductivity, divided by renormalization factor $Z$ of sector $B$ and plotted against doping.
}
\label{massenhancementB}
\end{figure}

The solid points in the upper panel of Fig.~\ref{massenhancementB} show the value of $Z$ for the sector $B$ containing the zone-diagonal point $(\pi/2,\pi/2)$ and the sector $C$ containing the zone-face point $(0,\pi)$ for dopings for which the sectors are Fermi-liquid-like. The open symbols show the mathematically defined values of $Z$ in the regime where it has no physical meaning because the regime is not Fermi-liquid-like. For dopings in the Fermi-liquid regime the $Z$ in sector $B$ is linear in $x$ but extrapolates to a small non-zero value at $x=0$. This is approximately but not exactly the behavior $Z \sim x$ expected in a doped Mott insulator. The lower panel of Fig.~\ref{massenhancementB} shows that the doping dependence of the low frequency optical conductivity weight is essentially the same as that of the nodal-sector $Z$.

\section{Summary \label{Conclusion}}

In the $8$-site DCA approximation to the solution of the two dimensional Hubbard model, the doping-driven Mott transition occurs in an orbitally selective manner. As doping is reduced a pseudogap opens in  the region of momentum space centered at $(0,\pi)$ (and $(\pi,0)$) while the sector containing the zone-diagonal remains ungapped down to much lower dopings. In this paper we have explored some of the implications of this pseudogap for observables. We find that the gap is apparently tied to the Fermi level, fills in rather than closes with increasing temperature, and produces many of the qualitative features observed in experiment.

\begin{figure}[thbp]
\includegraphics[angle=-0,width=0.95\columnwidth]{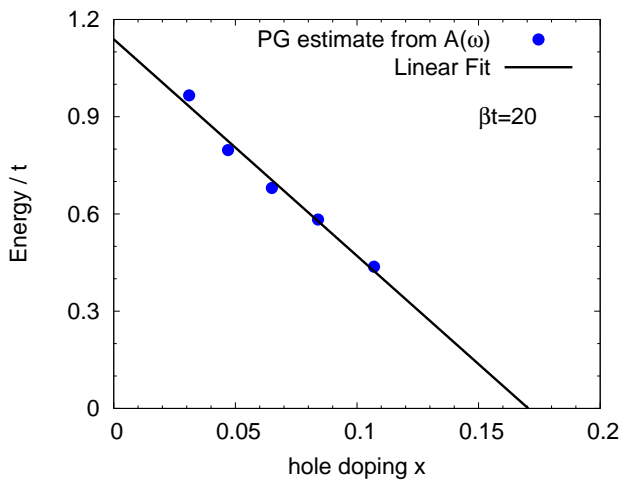}
\caption{Points: pseudogap size determined from electron spectral function as a function of hole doping $x$ for $U=7t$ and inverse temperature $\beta=20/t\approx200\ \text{K}$. Line: linear fit to results.}
\label{pseudogapvsdoping}
\end{figure}

Fig.~\ref{pseudogapvsdoping} plots the doping dependence of the pseudogap magnitude as defined from the peak to peak separation in the electron spectral function. As is seen in experiment \cite{Hufner08} the pseudogap magnitude is approximately linear in doping. The extrapolation to $x=0$ (see Fig.~\ref{pseudogapvsdoping}) gives $\lim_{x\rightarrow0}2\Delta_{PG}(x)=1.1t$ rather smaller than the $x=0$ sector $C$ gap $\sim 2.6t$ characteristic of the half filled insulator. The pseudogap is thus a new phenomenon, not a remnant of the gap occurring at half filling.

For $x>0.11$, a gap is not visible in our calculated spectral functions; however, the linear extrapolation of the points in Fig.~\ref{pseudogapvsdoping} indicates that for $U=7t$ studied here the critical value at which the gap would close is $x\sim0.17$. One possible inference is that the temperatures $T\geq t/20$ accessible to us are too high to enable the gap to be seen, and that calculations at lower temperatures would reveal a gap in the range $0.11<x<0.17$. 
Indeed, in experiment pseudogap effects are visible at dopings $> 0.11$ but only at temperatures $< 200\ \text{K}$ (see e.~g. Ref.~\onlinecite{Damascelli03}, Fig.~63 or Ref.~\onlinecite{Yu08}), i.e. less than the lowest temperature accessed in this study. Extending our results to lower temperatures is therefore of interest.
It is also interesting to note that this larger doping is comparable with the value $x=0.15$ at which a quantum critical point was reported from analysis of the electron lifetime in a model with $U=6t$.\cite{Vidhyadhiraja09}

Fig.~\ref{pseudogapvsdoping} should however be interpreted with caution. At low dopings (see e.~g. the $x=0.05$ results in Fig.~\ref{spectral}) the low $T$ limit of the spectral function exhibits a clear gap (region of vanishing density of states) and the peak to peak distance plotted in Fig.~\ref{pseudogapvsdoping} corresponds well to the gap edge. As doping is increased the gap fills in and it is less clear whether as $T\rightarrow0$ the density of states would develop a true gap, or whether for $x \ge0.1$ the peaks represent the boundary of a region with reduced, but non-vanishing density of states. If the second possibility occurs then $x\sim0.1$ is a critical doping separating a low doping region where sector $C$ is gapped from an intermediate region where sector $C$ has a non-zero, but possibly suppressed Fermi level density of states. This latter possibility is suggested by the imaginary axis analysis of previous papers \cite{Werner09,Gull09} which defined the critical doping for the orbital selective transition in terms of the chemical potential at which carriers were first added to sector $C$. This chemical potential value implies a critical $x\sim 0.1$.

A closely related question concerns the possible formation of Fermi arcs. A natural interpretation of the results is that as doping is decreased the pseudogap first forms at $(0,\pi)/(\pi,0)$ and with further decrease of doping an increasing portion of the Fermi surface is gapped, leaving a ``Fermi arc'' whose width is doping dependent. In this interpretation the sector $B$ finding of a scattering rate which becomes large as doping is decreased would represent an average over a region (increasing as doping is decreased) where the Fermi surface is gapped and a Fermi arc region (decreasing as doping is decreased) with good quasiparticles. Analysis of this possibility requires a finer momentum resolution than is presently available to us.

While differences remain on the quantitative level between calculation and experiment and indeed between calculations performed on different clusters, the results indicate clearly that the Hubbard model at intermediate correlations and low dopings does exhibit a pseudogap with many of the features exhibited by the experimentally-defined high energy pseudogap. A transition to a phase with long ranged order is not necessary to produce the effect. The calculation does not reproduce many of the lower energy anomalies which may be associated with onset of significant superconducting, nematic, or orbital current order, perhaps because these are long wavelength effects beyond the scale provided by the cluster sizes or because they involve physics beyond the one-band model we are presently able to study. 

{\it Acknowledgements} We thank C. Bernhard and A. Dubroka for discussion of the interplane conductivity  and R. Hackl for discussion of Raman scattering.  We acknowledge support from NSF-DMR-0705847.  AJM and EG also acknowledge partial support from the  National Science Foundation under Grant No. PHY05-51164. QMC calculations have been performed using a code based on the ALPS\cite{ALPS} library on the Brutus cluster at ETH Z\"{u}rich. A portion of this research was conducted at the Center for Nanophase Materials Sciences, which is sponsored at Oak Ridge National Laboratory by the Division of Scientific User Facilities, U.S. Department of Energy.

\appendix
\section{Determination of Fermi-liquid regime} \label{detflregime}
\begin{figure}[htbp]
\begin{center}
\includegraphics[angle=-0,width=1.1\columnwidth]{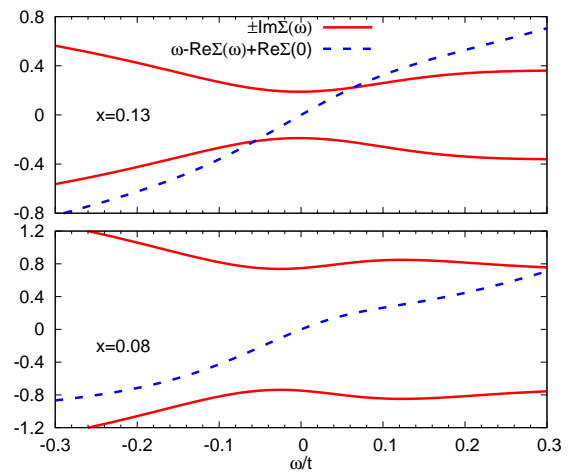}
\end{center}
\caption{Imaginary part of the $B$-sector self-energy (solid line, red on-line) and real part of the $B$-sector self-energy (dashed line, blue on-line) at dopings $x=0.13$ (upper panel) and $x=0.08$ (lower panel), as a function of frequency. Note the different $y$ axis scales.}
\label{fermiliquid}
\end{figure} 
In this appendix we provide details of the analysis we use to determine whether the system is in a Fermi-liquid regime. 
Representative results for $\text{Re}\Sigma(\omega)$ and $\text{Im}\Sigma(\omega)$ are shown in Fig.~\ref{fermiliquid}.
In a  Fermi-liquid, the real part of the self-energy is linear in frequency (at low frequency), and the imaginary part is not too big. 
To formulate a quantitative criterion we first determine the range around $\omega=0$ over which $\text{Re}\Sigma(\omega)$ is linear. This range is bounded at the lower end by a frequency $\omega_L$ and at the higher end by $\omega_H$. For the $x=0.13$ data we have $\omega_L =  -0.08$, $\omega_H = 0.08$, for $x=0.08$ we have $\omega_L = -0.08$, $\omega_H = 0.02$. Then we compute the change in quasiparticle energy $(\omega_H - \Sigma(\omega_H)) - (\omega_L - \Sigma(\omega_L))$ (finding $0.56596$ for $x=0.13$ and $0.42048$ for $x=0.08$). Finally we compare the result to $-2\text{Im}\Sigma(\omega=0)$ which changes considerably between dopings. From this comparison we see that sector $B$ at $x=0.13$ is well within the Fermi-liquid region while $x=0.08$ is on the border.

\section{Real Part of Self Energies}
\begin{figure}[htbp]
\centering
\includegraphics[angle=-0,width=0.95\columnwidth]{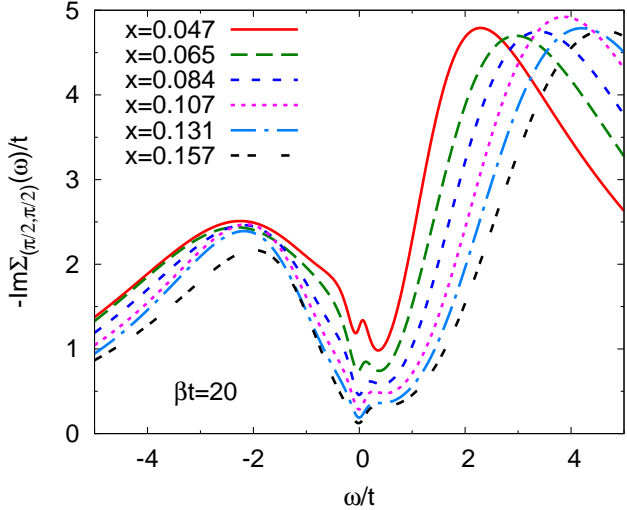}
\caption{Imaginary part of the self-energy in sector $B$ as a function of frequency, for dopings indicated, corresponding to the real part in Fig.~\ref{resigma}.}
\label{ImSigB}
\end{figure}

\begin{figure}[H]
\centering
\includegraphics[angle=-0,width=0.95\columnwidth]{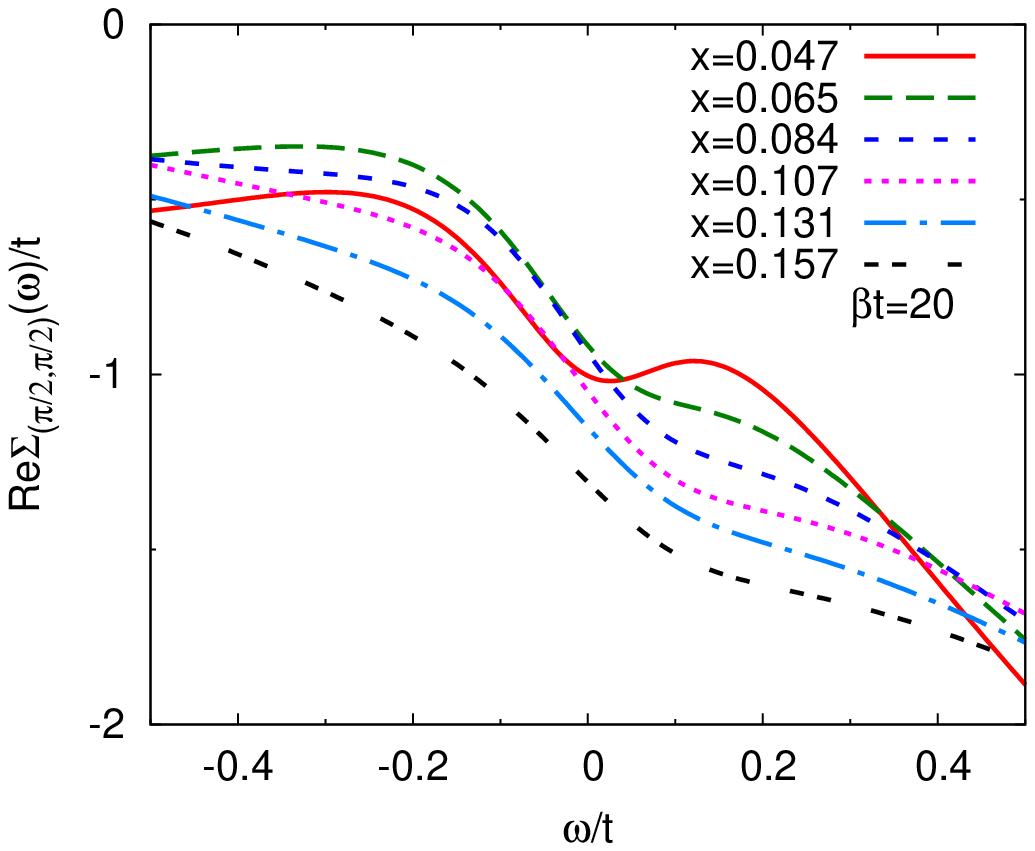}
\includegraphics[angle=-0,width=0.95\columnwidth]{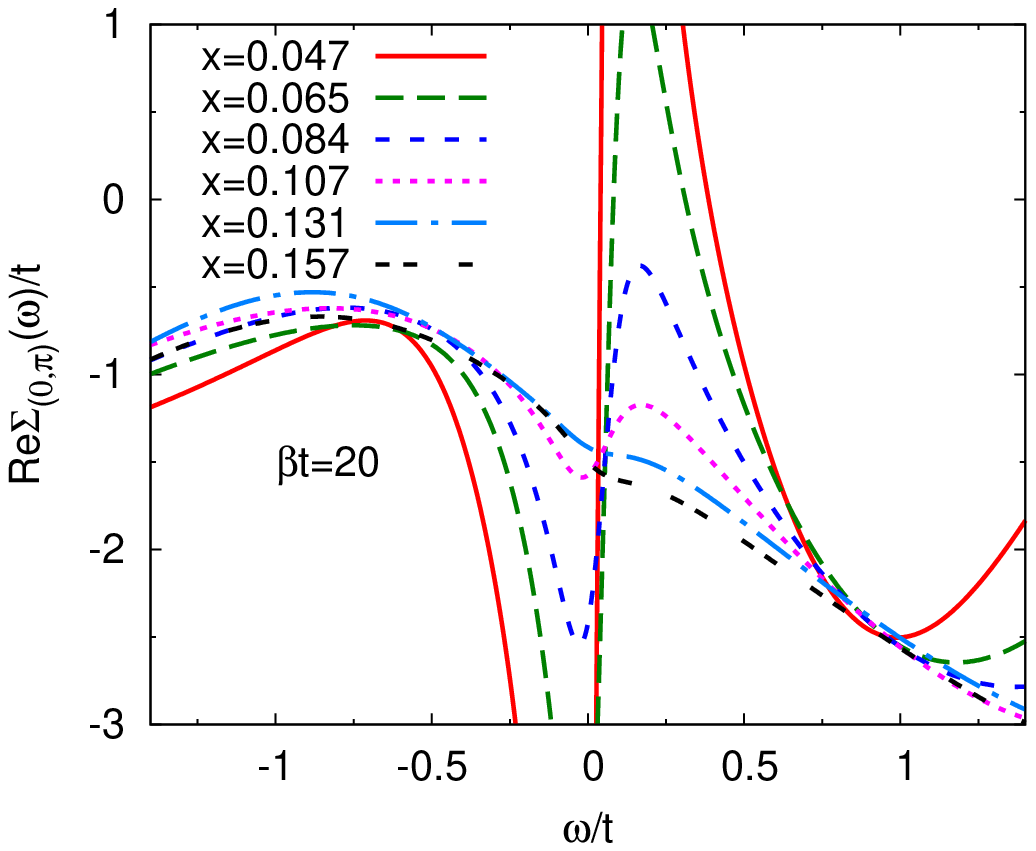}
\caption{Real part of the self-energy in sector $B$ (upper panel) and $C$ (lower panel) as a function of frequency, for dopings indicated, corresponding to the imaginary parts of the self-energies displayed in Fig.~\ref{ImSigC} and Fig.~\ref{ImSigB}.}
\label{resigma}
\end{figure}

For completeness we show in this appendix the self-energies not discussed in the text. Fig.~\ref{ImSigB} presents the imaginary part of the self-energy in sector $B$ containing the $(\pm \pi/2,\pm\pi/2)$ point. The real parts of the self-energies in sector $B$ and sector $C$ over an intermeidate frequency range are shown in Fig.~\ref{resigma}.

\bibliography{pseudogap.bbl}

\begin{thebibliography}{82}
\expandafter\ifx\csname natexlab\endcsname\relax\def\natexlab#1{#1}\fi
\expandafter\ifx\csname bibnamefont\endcsname\relax
  \def\bibnamefont#1{#1}\fi
\expandafter\ifx\csname bibfnamefont\endcsname\relax
  \def\bibfnamefont#1{#1}\fi
\expandafter\ifx\csname citenamefont\endcsname\relax
  \def\citenamefont#1{#1}\fi
\expandafter\ifx\csname url\endcsname\relax
  \def\url#1{\texttt{#1}}\fi
\expandafter\ifx\csname urlprefix\endcsname\relax\def\urlprefix{URL }\fi
\providecommand{\bibinfo}[2]{#2}
\providecommand{\eprint}[2][]{\url{#2}}

\bibitem[{\citenamefont{Warren et~al.}(1989)\citenamefont{Warren, Walstedt,
  Brennert, Cava, Tycko, Bell, and Dabbagh}}]{Warren89}
\bibinfo{author}{\bibfnamefont{W.~W.} \bibnamefont{Warren}},
  \bibinfo{author}{\bibfnamefont{R.~E.} \bibnamefont{Walstedt}},
  \bibinfo{author}{\bibfnamefont{G.~F.} \bibnamefont{Brennert}},
  \bibinfo{author}{\bibfnamefont{R.~J.} \bibnamefont{Cava}},
  \bibinfo{author}{\bibfnamefont{R.}~\bibnamefont{Tycko}},
  \bibinfo{author}{\bibfnamefont{R.~F.} \bibnamefont{Bell}}, \bibnamefont{and}
  \bibinfo{author}{\bibfnamefont{G.}~\bibnamefont{Dabbagh}},
  \bibinfo{journal}{Phys. Rev. Lett.} \textbf{\bibinfo{volume}{62}},
  \bibinfo{pages}{1193} (\bibinfo{year}{1989}).

\bibitem[{\citenamefont{Alloul et~al.}(1989)\citenamefont{Alloul, Ohno, and
  Mendels}}]{Alloul89}
\bibinfo{author}{\bibfnamefont{H.}~\bibnamefont{Alloul}},
  \bibinfo{author}{\bibfnamefont{T.}~\bibnamefont{Ohno}}, \bibnamefont{and}
  \bibinfo{author}{\bibfnamefont{P.}~\bibnamefont{Mendels}},
  \bibinfo{journal}{Phys. Rev. Lett.} \textbf{\bibinfo{volume}{63}},
  \bibinfo{pages}{1700} (\bibinfo{year}{1989}).

\bibitem[{\citenamefont{Ito et~al.}(1993)\citenamefont{Ito, Takenaka, and
  Uchida}}]{Ito93}
\bibinfo{author}{\bibfnamefont{T.}~\bibnamefont{Ito}},
  \bibinfo{author}{\bibfnamefont{K.}~\bibnamefont{Takenaka}}, \bibnamefont{and}
  \bibinfo{author}{\bibfnamefont{S.}~\bibnamefont{Uchida}},
  \bibinfo{journal}{Phys. Rev. Lett.} \textbf{\bibinfo{volume}{70}},
  \bibinfo{pages}{3995} (\bibinfo{year}{1993}).

\bibitem[{\citenamefont{Loeser et~al.}(1996)\citenamefont{Loeser, Shen, Dessau,
  Marshall, Park, Fournier, and Kapitulnik}}]{Loeser96}
\bibinfo{author}{\bibfnamefont{A.~G.} \bibnamefont{Loeser}},
  \bibinfo{author}{\bibfnamefont{Z.-X.} \bibnamefont{Shen}},
  \bibinfo{author}{\bibfnamefont{D.~S.} \bibnamefont{Dessau}},
  \bibinfo{author}{\bibfnamefont{D.~S.} \bibnamefont{Marshall}},
  \bibinfo{author}{\bibfnamefont{C.~H.} \bibnamefont{Park}},
  \bibinfo{author}{\bibfnamefont{P.}~\bibnamefont{Fournier}}, \bibnamefont{and}
  \bibinfo{author}{\bibfnamefont{A.}~\bibnamefont{Kapitulnik}},
  \bibinfo{journal}{Science} \textbf{\bibinfo{volume}{273}},
  \bibinfo{pages}{325} (\bibinfo{year}{1996}).

\bibitem[{\citenamefont{Ding et~al.}(1996)\citenamefont{Ding, Yokoya,
  Campuzano, Takahashi, Randeria, Norman, Mochiku, Kadowaki, and
  Giapintzakis}}]{Ding96}
\bibinfo{author}{\bibfnamefont{H.}~\bibnamefont{Ding}},
  \bibinfo{author}{\bibfnamefont{T.}~\bibnamefont{Yokoya}},
  \bibinfo{author}{\bibfnamefont{J.~C.} \bibnamefont{Campuzano}},
  \bibinfo{author}{\bibfnamefont{T.}~\bibnamefont{Takahashi}},
  \bibinfo{author}{\bibfnamefont{M.}~\bibnamefont{Randeria}},
  \bibinfo{author}{\bibfnamefont{M.~R.} \bibnamefont{Norman}},
  \bibinfo{author}{\bibfnamefont{T.}~\bibnamefont{Mochiku}},
  \bibinfo{author}{\bibfnamefont{K.}~\bibnamefont{Kadowaki}}, \bibnamefont{and}
  \bibinfo{author}{\bibfnamefont{J.}~\bibnamefont{Giapintzakis}},
  \bibinfo{journal}{Nature} \textbf{\bibinfo{volume}{382}}, \bibinfo{pages}{51
  } (\bibinfo{year}{1996}).

\bibitem[{\citenamefont{Homes et~al.}(1993)\citenamefont{Homes, Timusk, Liang,
  Bonn, and Hardy}}]{Homes93}
\bibinfo{author}{\bibfnamefont{C.~C.} \bibnamefont{Homes}},
  \bibinfo{author}{\bibfnamefont{T.}~\bibnamefont{Timusk}},
  \bibinfo{author}{\bibfnamefont{R.}~\bibnamefont{Liang}},
  \bibinfo{author}{\bibfnamefont{D.~A.} \bibnamefont{Bonn}}, \bibnamefont{and}
  \bibinfo{author}{\bibfnamefont{W.~N.} \bibnamefont{Hardy}},
  \bibinfo{journal}{Phys. Rev. Lett.} \textbf{\bibinfo{volume}{71}},
  \bibinfo{pages}{1645} (\bibinfo{year}{1993}).

\bibitem[{\citenamefont{Tajima et~al.}(1997)\citenamefont{Tajima, Sch\"utzmann,
  Miyamoto, Terasaki, Sato, and Hauff}}]{Tajima97}
\bibinfo{author}{\bibfnamefont{S.}~\bibnamefont{Tajima}},
  \bibinfo{author}{\bibfnamefont{J.}~\bibnamefont{Sch\"utzmann}},
  \bibinfo{author}{\bibfnamefont{S.}~\bibnamefont{Miyamoto}},
  \bibinfo{author}{\bibfnamefont{I.}~\bibnamefont{Terasaki}},
  \bibinfo{author}{\bibfnamefont{Y.}~\bibnamefont{Sato}}, \bibnamefont{and}
  \bibinfo{author}{\bibfnamefont{R.}~\bibnamefont{Hauff}},
  \bibinfo{journal}{Phys. Rev. B} \textbf{\bibinfo{volume}{55}},
  \bibinfo{pages}{6051} (\bibinfo{year}{1997}).

\bibitem[{\citenamefont{Nemetschek et~al.}(1997)\citenamefont{Nemetschek, Opel,
  Hoffmann, M\"uller, Hackl, Berger, Forr\'o, Erb, and Walker}}]{Nemetschek97}
\bibinfo{author}{\bibfnamefont{R.}~\bibnamefont{Nemetschek}},
  \bibinfo{author}{\bibfnamefont{M.}~\bibnamefont{Opel}},
  \bibinfo{author}{\bibfnamefont{C.}~\bibnamefont{Hoffmann}},
  \bibinfo{author}{\bibfnamefont{P.~F.} \bibnamefont{M\"uller}},
  \bibinfo{author}{\bibfnamefont{R.}~\bibnamefont{Hackl}},
  \bibinfo{author}{\bibfnamefont{H.}~\bibnamefont{Berger}},
  \bibinfo{author}{\bibfnamefont{L.}~\bibnamefont{Forr\'o}},
  \bibinfo{author}{\bibfnamefont{A.}~\bibnamefont{Erb}}, \bibnamefont{and}
  \bibinfo{author}{\bibfnamefont{E.}~\bibnamefont{Walker}},
  \bibinfo{journal}{Phys. Rev. Lett.} \textbf{\bibinfo{volume}{78}},
  \bibinfo{pages}{4837} (\bibinfo{year}{1997}).

\bibitem[{\citenamefont{Chen et~al.}(1997)\citenamefont{Chen, Naeini, Hewitt,
  Irwin, Liang, and Hardy}}]{Chen97}
\bibinfo{author}{\bibfnamefont{X.~K.} \bibnamefont{Chen}},
  \bibinfo{author}{\bibfnamefont{J.~G.} \bibnamefont{Naeini}},
  \bibinfo{author}{\bibfnamefont{K.~C.} \bibnamefont{Hewitt}},
  \bibinfo{author}{\bibfnamefont{J.~C.} \bibnamefont{Irwin}},
  \bibinfo{author}{\bibfnamefont{R.}~\bibnamefont{Liang}}, \bibnamefont{and}
  \bibinfo{author}{\bibfnamefont{W.~N.} \bibnamefont{Hardy}},
  \bibinfo{journal}{Phys. Rev. B} \textbf{\bibinfo{volume}{56}},
  \bibinfo{pages}{R513} (\bibinfo{year}{1997}).

\bibitem[{\citenamefont{Renner et~al.}(1998)\citenamefont{Renner, Revaz,
  Genoud, Kadowaki, and Fischer}}]{Renner98}
\bibinfo{author}{\bibfnamefont{C.}~\bibnamefont{Renner}},
  \bibinfo{author}{\bibfnamefont{B.}~\bibnamefont{Revaz}},
  \bibinfo{author}{\bibfnamefont{J.-Y.} \bibnamefont{Genoud}},
  \bibinfo{author}{\bibfnamefont{K.}~\bibnamefont{Kadowaki}}, \bibnamefont{and}
  \bibinfo{author}{\bibfnamefont{O.}~\bibnamefont{Fischer}},
  \bibinfo{journal}{Phys. Rev. Lett.} \textbf{\bibinfo{volume}{80}},
  \bibinfo{pages}{149} (\bibinfo{year}{1998}).

\bibitem[{\citenamefont{Orenstein et~al.}(1990)\citenamefont{Orenstein, Thomas,
  Millis, Cooper, Rapkine, Timusk, Schneemeyer, and Waszczak}}]{Orenstein91}
\bibinfo{author}{\bibfnamefont{J.}~\bibnamefont{Orenstein}},
  \bibinfo{author}{\bibfnamefont{G.~A.} \bibnamefont{Thomas}},
  \bibinfo{author}{\bibfnamefont{A.~J.} \bibnamefont{Millis}},
  \bibinfo{author}{\bibfnamefont{S.~L.} \bibnamefont{Cooper}},
  \bibinfo{author}{\bibfnamefont{D.~H.} \bibnamefont{Rapkine}},
  \bibinfo{author}{\bibfnamefont{T.}~\bibnamefont{Timusk}},
  \bibinfo{author}{\bibfnamefont{L.~F.} \bibnamefont{Schneemeyer}},
  \bibnamefont{and} \bibinfo{author}{\bibfnamefont{J.~V.}
  \bibnamefont{Waszczak}}, \bibinfo{journal}{Phys. Rev. B}
  \textbf{\bibinfo{volume}{42}}, \bibinfo{pages}{6342} (\bibinfo{year}{1990}).

\bibitem[{\citenamefont{Basov and Timusk}(2005)}]{Basov06}
\bibinfo{author}{\bibfnamefont{D.~N.} \bibnamefont{Basov}} \bibnamefont{and}
  \bibinfo{author}{\bibfnamefont{T.}~\bibnamefont{Timusk}},
  \bibinfo{journal}{Rev. Mod. Phys.} \textbf{\bibinfo{volume}{77}},
  \bibinfo{pages}{721} (\bibinfo{year}{2005}).

\bibitem[{\citenamefont{Damascelli et~al.}(2003)\citenamefont{Damascelli,
  Hussain, and Shen}}]{Damascelli03}
\bibinfo{author}{\bibfnamefont{A.}~\bibnamefont{Damascelli}},
  \bibinfo{author}{\bibfnamefont{Z.}~\bibnamefont{Hussain}}, \bibnamefont{and}
  \bibinfo{author}{\bibfnamefont{Z.-X.} \bibnamefont{Shen}},
  \bibinfo{journal}{Rev. Mod. Phys.} \textbf{\bibinfo{volume}{75}},
  \bibinfo{pages}{473} (\bibinfo{year}{2003}).

\bibitem[{\citenamefont{Schmalian et~al.}(1998)\citenamefont{Schmalian, Pines,
  and Stojkovi\ifmmode~\acute{c}\else \'{c}\fi{}}}]{Schmalian98}
\bibinfo{author}{\bibfnamefont{J.}~\bibnamefont{Schmalian}},
  \bibinfo{author}{\bibfnamefont{D.}~\bibnamefont{Pines}}, \bibnamefont{and}
  \bibinfo{author}{\bibfnamefont{B.}~\bibnamefont{Stojkovi\ifmmode~\acute{c}\e%
lse \'{c}\fi{}}}, \bibinfo{journal}{Phys. Rev. Lett.}
  \textbf{\bibinfo{volume}{80}}, \bibinfo{pages}{3839} (\bibinfo{year}{1998}).

\bibitem[{\citenamefont{Kivelson et~al.}(1998)\citenamefont{Kivelson, Fradkin,
  and Emery}}]{Kivelson98}
\bibinfo{author}{\bibfnamefont{S.~A.} \bibnamefont{Kivelson}},
  \bibinfo{author}{\bibfnamefont{E.}~\bibnamefont{Fradkin}}, \bibnamefont{and}
  \bibinfo{author}{\bibfnamefont{V.~J.} \bibnamefont{Emery}},
  \bibinfo{journal}{Nature} \textbf{\bibinfo{volume}{393}},
  \bibinfo{pages}{550} (\bibinfo{year}{1998}).

\bibitem[{\citenamefont{Varma}(2006)}]{Varma06}
\bibinfo{author}{\bibfnamefont{C.~M.} \bibnamefont{Varma}},
  \bibinfo{journal}{Phys. Rev. B} \textbf{\bibinfo{volume}{73}},
  \bibinfo{pages}{155113} (\bibinfo{year}{2006}).

\bibitem[{\citenamefont{Tranquada et~al.}(1995)\citenamefont{Tranquada,
  Lorenzo, Buttrey, and Sachan}}]{Tranquada95}
\bibinfo{author}{\bibfnamefont{J.~M.} \bibnamefont{Tranquada}},
  \bibinfo{author}{\bibfnamefont{J.~E.} \bibnamefont{Lorenzo}},
  \bibinfo{author}{\bibfnamefont{D.~J.} \bibnamefont{Buttrey}},
  \bibnamefont{and} \bibinfo{author}{\bibfnamefont{V.}~\bibnamefont{Sachan}},
  \bibinfo{journal}{Phys. Rev. B} \textbf{\bibinfo{volume}{52}},
  \bibinfo{pages}{3581} (\bibinfo{year}{1995}).

\bibitem[{\citenamefont{Fauqu\'e et~al.}(2006)\citenamefont{Fauqu\'e, Sidis,
  Hinkov, Pailh\`es, Lin, Chaud, and Bourges}}]{Fauque06}
\bibinfo{author}{\bibfnamefont{B.}~\bibnamefont{Fauqu\'e}},
  \bibinfo{author}{\bibfnamefont{Y.}~\bibnamefont{Sidis}},
  \bibinfo{author}{\bibfnamefont{V.}~\bibnamefont{Hinkov}},
  \bibinfo{author}{\bibfnamefont{S.}~\bibnamefont{Pailh\`es}},
  \bibinfo{author}{\bibfnamefont{C.~T.} \bibnamefont{Lin}},
  \bibinfo{author}{\bibfnamefont{X.}~\bibnamefont{Chaud}}, \bibnamefont{and}
  \bibinfo{author}{\bibfnamefont{P.}~\bibnamefont{Bourges}},
  \bibinfo{journal}{Phys. Rev. Lett.} \textbf{\bibinfo{volume}{96}},
  \bibinfo{pages}{197001} (\bibinfo{year}{2006}).

\bibitem[{\citenamefont{Sonier et~al.}(2009)\citenamefont{Sonier, Pacradouni,
  Sabok-Sayr, Hardy, Bonn, Liang, and Mook}}]{Sonier09}
\bibinfo{author}{\bibfnamefont{J.~E.} \bibnamefont{Sonier}},
  \bibinfo{author}{\bibfnamefont{V.}~\bibnamefont{Pacradouni}},
  \bibinfo{author}{\bibfnamefont{S.~A.} \bibnamefont{Sabok-Sayr}},
  \bibinfo{author}{\bibfnamefont{W.~N.} \bibnamefont{Hardy}},
  \bibinfo{author}{\bibfnamefont{D.~A.} \bibnamefont{Bonn}},
  \bibinfo{author}{\bibfnamefont{R.}~\bibnamefont{Liang}}, \bibnamefont{and}
  \bibinfo{author}{\bibfnamefont{H.~A.} \bibnamefont{Mook}},
  \bibinfo{journal}{Phys. Rev. Lett.} \textbf{\bibinfo{volume}{103}},
  \bibinfo{pages}{167002} (\bibinfo{year}{2009}).

\bibitem[{\citenamefont{Millis and Monien}(1993)}]{Millis93}
\bibinfo{author}{\bibfnamefont{A.~J.} \bibnamefont{Millis}} \bibnamefont{and}
  \bibinfo{author}{\bibfnamefont{H.}~\bibnamefont{Monien}},
  \bibinfo{journal}{Phys. Rev. Lett.} \textbf{\bibinfo{volume}{70}},
  \bibinfo{pages}{2810} (\bibinfo{year}{1993}).

\bibitem[{\citenamefont{Vilk and Tremblay}(1997)}]{Vilk97}
\bibinfo{author}{\bibfnamefont{Y.}~\bibnamefont{Vilk}} \bibnamefont{and}
  \bibinfo{author}{\bibfnamefont{A.-M.} \bibnamefont{Tremblay}},
  \bibinfo{journal}{J. Phys. I France} \textbf{\bibinfo{volume}{7}},
  \bibinfo{pages}{1309} (\bibinfo{year}{1997}).

\bibitem[{\citenamefont{Abanov et~al.}(2001{\natexlab{a}})\citenamefont{Abanov,
  Chubukov, and Schmalian}}]{Abanov01}
\bibinfo{author}{\bibfnamefont{A.}~\bibnamefont{Abanov}},
  \bibinfo{author}{\bibfnamefont{A.~V.} \bibnamefont{Chubukov}},
  \bibnamefont{and}
  \bibinfo{author}{\bibfnamefont{J.}~\bibnamefont{Schmalian}},
  \bibinfo{journal}{EPL} \textbf{\bibinfo{volume}{55}}, \bibinfo{pages}{369}
  (\bibinfo{year}{2001}{\natexlab{a}}).

\bibitem[{\citenamefont{Emery and Kivelson}(1995)}]{Emery95}
\bibinfo{author}{\bibfnamefont{V.~J.} \bibnamefont{Emery}} \bibnamefont{and}
  \bibinfo{author}{\bibfnamefont{S.~A.} \bibnamefont{Kivelson}},
  \bibinfo{journal}{Nature} \textbf{\bibinfo{volume}{374}}, \bibinfo{pages}{434
  } (\bibinfo{year}{1995}).

\bibitem[{\citenamefont{Wang et~al.}(2002)\citenamefont{Wang, Ong, Xu,
  Kakeshita, Uchida, Bonn, Liang, and Hardy}}]{Wang02}
\bibinfo{author}{\bibfnamefont{Y.}~\bibnamefont{Wang}},
  \bibinfo{author}{\bibfnamefont{N.~P.} \bibnamefont{Ong}},
  \bibinfo{author}{\bibfnamefont{Z.~A.} \bibnamefont{Xu}},
  \bibinfo{author}{\bibfnamefont{T.}~\bibnamefont{Kakeshita}},
  \bibinfo{author}{\bibfnamefont{S.}~\bibnamefont{Uchida}},
  \bibinfo{author}{\bibfnamefont{D.~A.} \bibnamefont{Bonn}},
  \bibinfo{author}{\bibfnamefont{R.}~\bibnamefont{Liang}}, \bibnamefont{and}
  \bibinfo{author}{\bibfnamefont{W.~N.} \bibnamefont{Hardy}},
  \bibinfo{journal}{Phys. Rev. Lett.} \textbf{\bibinfo{volume}{88}},
  \bibinfo{pages}{257003} (\bibinfo{year}{2002}).

\bibitem[{\citenamefont{Kotliar and Liu}(1988)}]{Kotliar88}
\bibinfo{author}{\bibfnamefont{G.}~\bibnamefont{Kotliar}} \bibnamefont{and}
  \bibinfo{author}{\bibfnamefont{J.}~\bibnamefont{Liu}},
  \bibinfo{journal}{Phys. Rev. B} \textbf{\bibinfo{volume}{38}},
  \bibinfo{pages}{5142} (\bibinfo{year}{1988}).

\bibitem[{\citenamefont{Lee and Nagaosa}(1992)}]{Lee92}
\bibinfo{author}{\bibfnamefont{P.~A.} \bibnamefont{Lee}} \bibnamefont{and}
  \bibinfo{author}{\bibfnamefont{N.}~\bibnamefont{Nagaosa}},
  \bibinfo{journal}{Phys. Rev. B} \textbf{\bibinfo{volume}{46}},
  \bibinfo{pages}{5621} (\bibinfo{year}{1992}).

\bibitem[{\citenamefont{Altshuler et~al.}(1996)\citenamefont{Altshuler, Ioffe,
  and Millis}}]{Altshuler96}
\bibinfo{author}{\bibfnamefont{B.~L.} \bibnamefont{Altshuler}},
  \bibinfo{author}{\bibfnamefont{L.~B.} \bibnamefont{Ioffe}}, \bibnamefont{and}
  \bibinfo{author}{\bibfnamefont{A.~J.} \bibnamefont{Millis}},
  \bibinfo{journal}{Phys. Rev. B} \textbf{\bibinfo{volume}{53}},
  \bibinfo{pages}{415} (\bibinfo{year}{1996}).

\bibitem[{\citenamefont{Lee et~al.}(1973)\citenamefont{Lee, Rice, and
  Anderson}}]{Lee73}
\bibinfo{author}{\bibfnamefont{P.~A.} \bibnamefont{Lee}},
  \bibinfo{author}{\bibfnamefont{T.~M.} \bibnamefont{Rice}}, \bibnamefont{and}
  \bibinfo{author}{\bibfnamefont{P.~W.} \bibnamefont{Anderson}},
  \bibinfo{journal}{Phys. Rev. Lett.} \textbf{\bibinfo{volume}{31}},
  \bibinfo{pages}{462} (\bibinfo{year}{1973}).

\bibitem[{\citenamefont{Dagotto and Rice}(1996)}]{Dagotto96}
\bibinfo{author}{\bibfnamefont{E.}~\bibnamefont{Dagotto}} \bibnamefont{and}
  \bibinfo{author}{\bibfnamefont{T.}~\bibnamefont{Rice}},
  \bibinfo{journal}{Science} \textbf{\bibinfo{volume}{271}},
  \bibinfo{pages}{618} (\bibinfo{year}{1996}).

\bibitem[{\citenamefont{Maier et~al.}(2005)\citenamefont{Maier, Jarrell,
  Pruschke, and Hettler}}]{Maier05}
\bibinfo{author}{\bibfnamefont{T.}~\bibnamefont{Maier}},
  \bibinfo{author}{\bibfnamefont{M.}~\bibnamefont{Jarrell}},
  \bibinfo{author}{\bibfnamefont{T.}~\bibnamefont{Pruschke}}, \bibnamefont{and}
  \bibinfo{author}{\bibfnamefont{M.~H.} \bibnamefont{Hettler}},
  \bibinfo{journal}{Rev. Mod. Phys.} \textbf{\bibinfo{volume}{77}},
  \bibinfo{eid}{1027} (\bibinfo{year}{2005}).

\bibitem[{\citenamefont{Huscroft et~al.}(2001)\citenamefont{Huscroft, Jarrell,
  Maier, Moukouri, and Tahvildarzadeh}}]{Huscroft01}
\bibinfo{author}{\bibfnamefont{C.}~\bibnamefont{Huscroft}},
  \bibinfo{author}{\bibfnamefont{M.}~\bibnamefont{Jarrell}},
  \bibinfo{author}{\bibfnamefont{T.}~\bibnamefont{Maier}},
  \bibinfo{author}{\bibfnamefont{S.}~\bibnamefont{Moukouri}}, \bibnamefont{and}
  \bibinfo{author}{\bibfnamefont{A.~N.} \bibnamefont{Tahvildarzadeh}},
  \bibinfo{journal}{Phys. Rev. Lett.} \textbf{\bibinfo{volume}{86}},
  \bibinfo{pages}{139} (\bibinfo{year}{2001}).

\bibitem[{\citenamefont{Parcollet et~al.}(2004)\citenamefont{Parcollet, Biroli,
  and Kotliar}}]{Parcollet04}
\bibinfo{author}{\bibfnamefont{O.}~\bibnamefont{Parcollet}},
  \bibinfo{author}{\bibfnamefont{G.}~\bibnamefont{Biroli}}, \bibnamefont{and}
  \bibinfo{author}{\bibfnamefont{G.}~\bibnamefont{Kotliar}},
  \bibinfo{journal}{Phys. Rev. Lett.} \textbf{\bibinfo{volume}{92}},
  \bibinfo{pages}{226402} (\bibinfo{year}{2004}).

\bibitem[{\citenamefont{Civelli et~al.}(2005)\citenamefont{Civelli, Capone,
  Kancharla, Parcollet, and Kotliar}}]{Civelli05}
\bibinfo{author}{\bibfnamefont{M.}~\bibnamefont{Civelli}},
  \bibinfo{author}{\bibfnamefont{M.}~\bibnamefont{Capone}},
  \bibinfo{author}{\bibfnamefont{S.~S.} \bibnamefont{Kancharla}},
  \bibinfo{author}{\bibfnamefont{O.}~\bibnamefont{Parcollet}},
  \bibnamefont{and} \bibinfo{author}{\bibfnamefont{G.}~\bibnamefont{Kotliar}},
  \bibinfo{journal}{Phys. Rev. Lett.} \textbf{\bibinfo{volume}{95}},
  \bibinfo{eid}{106402} (\bibinfo{year}{2005}).

\bibitem[{\citenamefont{Kyung et~al.}(2006)\citenamefont{Kyung, Kancharla,
  S\'{e}n\'{e}chal, Tremblay, Civelli, and Kotliar}}]{Kyung06}
\bibinfo{author}{\bibfnamefont{B.}~\bibnamefont{Kyung}},
  \bibinfo{author}{\bibfnamefont{S.~S.} \bibnamefont{Kancharla}},
  \bibinfo{author}{\bibfnamefont{D.}~\bibnamefont{S\'{e}n\'{e}chal}},
  \bibinfo{author}{\bibfnamefont{A.-M.~S.} \bibnamefont{Tremblay}},
  \bibinfo{author}{\bibfnamefont{M.}~\bibnamefont{Civelli}}, \bibnamefont{and}
  \bibinfo{author}{\bibfnamefont{G.}~\bibnamefont{Kotliar}},
  \bibinfo{journal}{Phys. Rev. B} \textbf{\bibinfo{volume}{73}},
  \bibinfo{eid}{165114} (\bibinfo{year}{2006}).

\bibitem[{\citenamefont{Stanescu and Kotliar}(2006)}]{Stanescu06}
\bibinfo{author}{\bibfnamefont{T.~D.} \bibnamefont{Stanescu}} \bibnamefont{and}
  \bibinfo{author}{\bibfnamefont{G.}~\bibnamefont{Kotliar}},
  \bibinfo{journal}{Phys. Rev. B} \textbf{\bibinfo{volume}{74}},
  \bibinfo{pages}{125110} (\bibinfo{year}{2006}).

\bibitem[{\citenamefont{Macridin et~al.}(2006)\citenamefont{Macridin, Jarrell,
  Maier, Kent, and D'Azevedo}}]{Macridin06}
\bibinfo{author}{\bibfnamefont{A.}~\bibnamefont{Macridin}},
  \bibinfo{author}{\bibfnamefont{M.}~\bibnamefont{Jarrell}},
  \bibinfo{author}{\bibfnamefont{T.}~\bibnamefont{Maier}},
  \bibinfo{author}{\bibfnamefont{P.~R.~C.} \bibnamefont{Kent}},
  \bibnamefont{and}
  \bibinfo{author}{\bibfnamefont{E.}~\bibnamefont{D'Azevedo}},
  \bibinfo{journal}{Phys. Rev. Lett.} \textbf{\bibinfo{volume}{97}},
  \bibinfo{pages}{036401} (\bibinfo{year}{2006}).

\bibitem[{\citenamefont{Zhang and Imada}(2007)}]{Zhang07}
\bibinfo{author}{\bibfnamefont{Y.~Z.} \bibnamefont{Zhang}} \bibnamefont{and}
  \bibinfo{author}{\bibfnamefont{M.}~\bibnamefont{Imada}},
  \bibinfo{journal}{Phys. Rev. B} \textbf{\bibinfo{volume}{76}},
  \bibinfo{pages}{045108} (\bibinfo{year}{2007}).

\bibitem[{\citenamefont{Civelli et~al.}(2008)\citenamefont{Civelli, Capone,
  Georges, Haule, Parcollet, Stanescu, and Kotliar}}]{Civelli08}
\bibinfo{author}{\bibfnamefont{M.}~\bibnamefont{Civelli}},
  \bibinfo{author}{\bibfnamefont{M.}~\bibnamefont{Capone}},
  \bibinfo{author}{\bibfnamefont{A.}~\bibnamefont{Georges}},
  \bibinfo{author}{\bibfnamefont{K.}~\bibnamefont{Haule}},
  \bibinfo{author}{\bibfnamefont{O.}~\bibnamefont{Parcollet}},
  \bibinfo{author}{\bibfnamefont{T.~D.} \bibnamefont{Stanescu}},
  \bibnamefont{and} \bibinfo{author}{\bibfnamefont{G.}~\bibnamefont{Kotliar}},
  \bibinfo{journal}{Phys. Rev. Lett.} \textbf{\bibinfo{volume}{100}},
  \bibinfo{pages}{046402} (\bibinfo{year}{2008}).

\bibitem[{\citenamefont{Gull et~al.}(2008{\natexlab{a}})\citenamefont{Gull,
  Werner, Wang, Troyer, and Millis}}]{Gull08}
\bibinfo{author}{\bibfnamefont{E.}~\bibnamefont{Gull}},
  \bibinfo{author}{\bibfnamefont{P.}~\bibnamefont{Werner}},
  \bibinfo{author}{\bibfnamefont{X.}~\bibnamefont{Wang}},
  \bibinfo{author}{\bibfnamefont{M.}~\bibnamefont{Troyer}}, \bibnamefont{and}
  \bibinfo{author}{\bibfnamefont{A.~J.} \bibnamefont{Millis}},
  \bibinfo{journal}{EPL} \textbf{\bibinfo{volume}{84}}, \bibinfo{pages}{37009}
  (\bibinfo{year}{2008}{\natexlab{a}}).

\bibitem[{\citenamefont{Park et~al.}(2008)\citenamefont{Park, Haule, and
  Kotliar}}]{Park08}
\bibinfo{author}{\bibfnamefont{H.}~\bibnamefont{Park}},
  \bibinfo{author}{\bibfnamefont{K.}~\bibnamefont{Haule}}, \bibnamefont{and}
  \bibinfo{author}{\bibfnamefont{G.}~\bibnamefont{Kotliar}},
  \bibinfo{journal}{Phys. Rev. Lett.} \textbf{\bibinfo{volume}{101}},
  \bibinfo{eid}{186403} (\bibinfo{year}{2008}).

\bibitem[{\citenamefont{Ferrero
  et~al.}(2009{\natexlab{a}})\citenamefont{Ferrero, Cornaglia, Leo, Parcollet,
  Kotliar, and Georges}}]{Ferrero09}
\bibinfo{author}{\bibfnamefont{M.}~\bibnamefont{Ferrero}},
  \bibinfo{author}{\bibfnamefont{P.~S.} \bibnamefont{Cornaglia}},
  \bibinfo{author}{\bibfnamefont{L.~D.} \bibnamefont{Leo}},
  \bibinfo{author}{\bibfnamefont{O.}~\bibnamefont{Parcollet}},
  \bibinfo{author}{\bibfnamefont{G.}~\bibnamefont{Kotliar}}, \bibnamefont{and}
  \bibinfo{author}{\bibfnamefont{A.}~\bibnamefont{Georges}},
  \bibinfo{journal}{EPL} \textbf{\bibinfo{volume}{85}}, \bibinfo{pages}{57009}
  (\bibinfo{year}{2009}{\natexlab{a}}).

\bibitem[{\citenamefont{Ferrero
  et~al.}(2009{\natexlab{b}})\citenamefont{Ferrero, Cornaglia, De~Leo,
  Parcollet, Kotliar, and Georges}}]{Ferrero09b}
\bibinfo{author}{\bibfnamefont{M.}~\bibnamefont{Ferrero}},
  \bibinfo{author}{\bibfnamefont{P.~S.} \bibnamefont{Cornaglia}},
  \bibinfo{author}{\bibfnamefont{L.}~\bibnamefont{De~Leo}},
  \bibinfo{author}{\bibfnamefont{O.}~\bibnamefont{Parcollet}},
  \bibinfo{author}{\bibfnamefont{G.}~\bibnamefont{Kotliar}}, \bibnamefont{and}
  \bibinfo{author}{\bibfnamefont{A.}~\bibnamefont{Georges}},
  \bibinfo{journal}{Phys. Rev. B} \textbf{\bibinfo{volume}{80}},
  \bibinfo{pages}{064501} (\bibinfo{year}{2009}{\natexlab{b}}).

\bibitem[{\citenamefont{Civelli}(2009)}]{Civelli09}
\bibinfo{author}{\bibfnamefont{M.}~\bibnamefont{Civelli}},
  \bibinfo{journal}{Phys. Rev. B} \textbf{\bibinfo{volume}{79}},
  \bibinfo{pages}{195113} (\bibinfo{year}{2009}).

\bibitem[{\citenamefont{Liebsch and Tong}(2009)}]{Liebsch09}
\bibinfo{author}{\bibfnamefont{A.}~\bibnamefont{Liebsch}} \bibnamefont{and}
  \bibinfo{author}{\bibfnamefont{N.-H.} \bibnamefont{Tong}},
  \bibinfo{journal}{Phys. Rev. B} \textbf{\bibinfo{volume}{80}},
  \bibinfo{pages}{165126} (\bibinfo{year}{2009}).

\bibitem[{\citenamefont{Sakai et~al.}(2009)\citenamefont{Sakai, Motome, and
  Imada}}]{Sakai09}
\bibinfo{author}{\bibfnamefont{S.}~\bibnamefont{Sakai}},
  \bibinfo{author}{\bibfnamefont{Y.}~\bibnamefont{Motome}}, \bibnamefont{and}
  \bibinfo{author}{\bibfnamefont{M.}~\bibnamefont{Imada}},
  \bibinfo{journal}{Phys. Rev. Lett.} \textbf{\bibinfo{volume}{102}},
  \bibinfo{pages}{056404} (\bibinfo{year}{2009}).

\bibitem[{\citenamefont{Sordi et~al.}(2010)\citenamefont{Sordi, Haule, and
  Tremblay}}]{Sordi10}
\bibinfo{author}{\bibfnamefont{G.}~\bibnamefont{Sordi}},
  \bibinfo{author}{\bibfnamefont{K.}~\bibnamefont{Haule}}, \bibnamefont{and}
  \bibinfo{author}{\bibfnamefont{A.-M.~S.} \bibnamefont{Tremblay}},
  \bibinfo{journal}{Phys. Rev. Lett.} \textbf{\bibinfo{volume}{104}},
  \bibinfo{pages}{226402} (\bibinfo{year}{2010}).

\bibitem[{\citenamefont{Sakai et~al.}(2010)\citenamefont{Sakai, Motome, and
  Imada}}]{Sakai10}
\bibinfo{author}{\bibfnamefont{S.}~\bibnamefont{Sakai}},
  \bibinfo{author}{\bibfnamefont{Y.}~\bibnamefont{Motome}}, \bibnamefont{and}
  \bibinfo{author}{\bibfnamefont{M.}~\bibnamefont{Imada}},
  \bibinfo{journal}{arXiv:1004.2569v1}  (\bibinfo{year}{2010}).

\bibitem[{\citenamefont{Ferrero et~al.}(2010)\citenamefont{Ferrero, Parcollet,
  Kotliar, and Georges}}]{Ferrero10}
\bibinfo{author}{\bibfnamefont{M.}~\bibnamefont{Ferrero}},
  \bibinfo{author}{\bibfnamefont{O.}~\bibnamefont{Parcollet}},
  \bibinfo{author}{\bibfnamefont{G.}~\bibnamefont{Kotliar}}, \bibnamefont{and}
  \bibinfo{author}{\bibfnamefont{A.}~\bibnamefont{Georges}},
  \bibinfo{journal}{arXiv:1001.5051v1}  (\bibinfo{year}{2010}).

\bibitem[{\citenamefont{Werner et~al.}(2009)\citenamefont{Werner, Gull,
  Parcollet, and Millis}}]{Werner09}
\bibinfo{author}{\bibfnamefont{P.}~\bibnamefont{Werner}},
  \bibinfo{author}{\bibfnamefont{E.}~\bibnamefont{Gull}},
  \bibinfo{author}{\bibfnamefont{O.}~\bibnamefont{Parcollet}},
  \bibnamefont{and} \bibinfo{author}{\bibfnamefont{A.~J.}
  \bibnamefont{Millis}}, \bibinfo{journal}{Phys. Rev. B}
  \textbf{\bibinfo{volume}{80}}, \bibinfo{eid}{045120} (\bibinfo{year}{2009}).

\bibitem[{\citenamefont{Gull et~al.}(2009)\citenamefont{Gull, Parcollet,
  Werner, and Millis}}]{Gull09}
\bibinfo{author}{\bibfnamefont{E.}~\bibnamefont{Gull}},
  \bibinfo{author}{\bibfnamefont{O.}~\bibnamefont{Parcollet}},
  \bibinfo{author}{\bibfnamefont{P.}~\bibnamefont{Werner}}, \bibnamefont{and}
  \bibinfo{author}{\bibfnamefont{A.~J.} \bibnamefont{Millis}},
  \bibinfo{journal}{Phys. Rev. B} \textbf{\bibinfo{volume}{80}},
  \bibinfo{pages}{245102} (\bibinfo{year}{2009}).

\bibitem[{\citenamefont{Andersen et~al.}(1994)\citenamefont{Andersen, Jepsen,
  Liechtenstein, and Mazin}}]{Andersen94}
\bibinfo{author}{\bibfnamefont{O.~K.} \bibnamefont{Andersen}},
  \bibinfo{author}{\bibfnamefont{O.}~\bibnamefont{Jepsen}},
  \bibinfo{author}{\bibfnamefont{A.~I.} \bibnamefont{Liechtenstein}},
  \bibnamefont{and} \bibinfo{author}{\bibfnamefont{I.~I.} \bibnamefont{Mazin}},
  \bibinfo{journal}{Phys. Rev. B} \textbf{\bibinfo{volume}{49}},
  \bibinfo{pages}{4145} (\bibinfo{year}{1994}).

\bibitem[{\citenamefont{Hettler et~al.}(1998)\citenamefont{Hettler,
  Tahvildar-Zadeh, Jarrell, Pruschke, and Krishnamurthy}}]{Hettler98}
\bibinfo{author}{\bibfnamefont{M.~H.} \bibnamefont{Hettler}},
  \bibinfo{author}{\bibfnamefont{A.~N.} \bibnamefont{Tahvildar-Zadeh}},
  \bibinfo{author}{\bibfnamefont{M.}~\bibnamefont{Jarrell}},
  \bibinfo{author}{\bibfnamefont{T.}~\bibnamefont{Pruschke}}, \bibnamefont{and}
  \bibinfo{author}{\bibfnamefont{H.~R.} \bibnamefont{Krishnamurthy}},
  \bibinfo{journal}{Phys. Rev. B} \textbf{\bibinfo{volume}{58}},
  \bibinfo{pages}{R7475} (\bibinfo{year}{1998}).

\bibitem[{\citenamefont{Gull et~al.}(2008{\natexlab{b}})\citenamefont{Gull,
  Werner, Parcollet, and Troyer}}]{Gull08_ctaux}
\bibinfo{author}{\bibfnamefont{E.}~\bibnamefont{Gull}},
  \bibinfo{author}{\bibfnamefont{P.}~\bibnamefont{Werner}},
  \bibinfo{author}{\bibfnamefont{O.}~\bibnamefont{Parcollet}},
  \bibnamefont{and} \bibinfo{author}{\bibfnamefont{M.}~\bibnamefont{Troyer}},
  \bibinfo{journal}{EPL} \textbf{\bibinfo{volume}{82}}, \bibinfo{pages}{57003}
  (\bibinfo{year}{2008}{\natexlab{b}}).

\bibitem[{\citenamefont{Wang et~al.}(2009)\citenamefont{Wang, Gull, de' Medici,
  Capone, and Millis}}]{Wang09}
\bibinfo{author}{\bibfnamefont{X.}~\bibnamefont{Wang}},
  \bibinfo{author}{\bibfnamefont{E.}~\bibnamefont{Gull}},
  \bibinfo{author}{\bibfnamefont{L.}~\bibnamefont{de' Medici}},
  \bibinfo{author}{\bibfnamefont{M.}~\bibnamefont{Capone}}, \bibnamefont{and}
  \bibinfo{author}{\bibfnamefont{A.~J.} \bibnamefont{Millis}},
  \bibinfo{journal}{Phys. Rev. B} \textbf{\bibinfo{volume}{80}},
  \bibinfo{pages}{045101} (\bibinfo{year}{2009}).

\bibitem[{\citenamefont{Jarrell and Gubernatis}(1996)}]{Jarrell96}
\bibinfo{author}{\bibfnamefont{M.}~\bibnamefont{Jarrell}} \bibnamefont{and}
  \bibinfo{author}{\bibfnamefont{J.~E.} \bibnamefont{Gubernatis}},
  \bibinfo{journal}{Physics Reports} \textbf{\bibinfo{volume}{269}},
  \bibinfo{pages}{133 } (\bibinfo{year}{1996}).

\bibitem[{\citenamefont{Rabani et~al.}(2002)\citenamefont{Rabani, Reichman,
  Krilov, and Berne}}]{Rabani02}
\bibinfo{author}{\bibfnamefont{E.}~\bibnamefont{Rabani}},
  \bibinfo{author}{\bibfnamefont{D.~R.} \bibnamefont{Reichman}},
  \bibinfo{author}{\bibfnamefont{G.}~\bibnamefont{Krilov}}, \bibnamefont{and}
  \bibinfo{author}{\bibfnamefont{B.~J.} \bibnamefont{Berne}},
  \bibinfo{journal}{Proc. Natl, Acad. Sci.} \textbf{\bibinfo{volume}{99}},
  \bibinfo{pages}{1129} (\bibinfo{year}{2002}).

\bibitem[{\citenamefont{Comanac et~al.}(2008)\citenamefont{Comanac, de' Medici,
  Capone, and Millis}}]{Comanac08}
\bibinfo{author}{\bibfnamefont{A.}~\bibnamefont{Comanac}},
  \bibinfo{author}{\bibfnamefont{L.}~\bibnamefont{de' Medici}},
  \bibinfo{author}{\bibfnamefont{M.}~\bibnamefont{Capone}}, \bibnamefont{and}
  \bibinfo{author}{\bibfnamefont{A.~J.} \bibnamefont{Millis}},
  \bibinfo{journal}{Nat Phys} \textbf{\bibinfo{volume}{4}},
  \bibinfo{pages}{287} (\bibinfo{year}{2008}).

\bibitem[{\citenamefont{Lin et~al.}(2009)\citenamefont{Lin, Gull, and
  Millis}}]{Lin09}
\bibinfo{author}{\bibfnamefont{N.}~\bibnamefont{Lin}},
  \bibinfo{author}{\bibfnamefont{E.}~\bibnamefont{Gull}}, \bibnamefont{and}
  \bibinfo{author}{\bibfnamefont{A.~J.} \bibnamefont{Millis}},
  \bibinfo{journal}{Phys. Rev. B} \textbf{\bibinfo{volume}{80}},
  \bibinfo{pages}{161105} (\bibinfo{year}{2009}).

\bibitem[{\citenamefont{Yu et~al.}(2008)\citenamefont{Yu, Munzar, Boris,
  Yordanov, Chaloupka, Wolf, Lin, Keimer, and Bernhard}}]{Yu08}
\bibinfo{author}{\bibfnamefont{L.}~\bibnamefont{Yu}},
  \bibinfo{author}{\bibfnamefont{D.}~\bibnamefont{Munzar}},
  \bibinfo{author}{\bibfnamefont{A.~V.} \bibnamefont{Boris}},
  \bibinfo{author}{\bibfnamefont{P.}~\bibnamefont{Yordanov}},
  \bibinfo{author}{\bibfnamefont{J.}~\bibnamefont{Chaloupka}},
  \bibinfo{author}{\bibfnamefont{T.}~\bibnamefont{Wolf}},
  \bibinfo{author}{\bibfnamefont{C.~T.} \bibnamefont{Lin}},
  \bibinfo{author}{\bibfnamefont{B.}~\bibnamefont{Keimer}}, \bibnamefont{and}
  \bibinfo{author}{\bibfnamefont{C.}~\bibnamefont{Bernhard}},
  \bibinfo{journal}{Phys. Rev. Lett.} \textbf{\bibinfo{volume}{100}},
  \bibinfo{pages}{177004} (\bibinfo{year}{2008}).

\bibitem[{\citenamefont{Duli\ifmmode~\acute{c}\else \'{c}\fi{}
  et~al.}(2001)\citenamefont{Duli\ifmmode~\acute{c}\else \'{c}\fi{}, Pimenov,
  van~der Marel, Broun, Kamal, Hardy, Tsvetkov, Sutjaha, Liang, Menovsky
  et~al.}}]{Dulic01}
\bibinfo{author}{\bibfnamefont{D.}~\bibnamefont{Duli\ifmmode~\acute{c}\else
  \'{c}\fi{}}}, \bibinfo{author}{\bibfnamefont{A.}~\bibnamefont{Pimenov}},
  \bibinfo{author}{\bibfnamefont{D.}~\bibnamefont{van~der Marel}},
  \bibinfo{author}{\bibfnamefont{D.~M.} \bibnamefont{Broun}},
  \bibinfo{author}{\bibfnamefont{S.}~\bibnamefont{Kamal}},
  \bibinfo{author}{\bibfnamefont{W.~N.} \bibnamefont{Hardy}},
  \bibinfo{author}{\bibfnamefont{A.~A.} \bibnamefont{Tsvetkov}},
  \bibinfo{author}{\bibfnamefont{I.~M.} \bibnamefont{Sutjaha}},
  \bibinfo{author}{\bibfnamefont{R.}~\bibnamefont{Liang}},
  \bibinfo{author}{\bibfnamefont{A.~A.} \bibnamefont{Menovsky}},
  \bibnamefont{et~al.}, \bibinfo{journal}{Phys. Rev. Lett.}
  \textbf{\bibinfo{volume}{86}}, \bibinfo{pages}{4144} (\bibinfo{year}{2001}).

\bibitem[{\citenamefont{Shah and Millis}(2001)}]{Shah01}
\bibinfo{author}{\bibfnamefont{N.}~\bibnamefont{Shah}} \bibnamefont{and}
  \bibinfo{author}{\bibfnamefont{A.~J.} \bibnamefont{Millis}},
  \bibinfo{journal}{Phys. Rev. B} \textbf{\bibinfo{volume}{65}},
  \bibinfo{pages}{024506} (\bibinfo{year}{2001}).

\bibitem[{\citenamefont{Devereaux and Hackl}(2007)}]{Devereaux07}
\bibinfo{author}{\bibfnamefont{T.~P.} \bibnamefont{Devereaux}}
  \bibnamefont{and} \bibinfo{author}{\bibfnamefont{R.}~\bibnamefont{Hackl}},
  \bibinfo{journal}{Rev. Mod. Phys.} \textbf{\bibinfo{volume}{79}},
  \bibinfo{pages}{175} (\bibinfo{year}{2007}).

\bibitem[{\citenamefont{de' Medici et~al.}(2008)\citenamefont{de' Medici,
  Georges, and Kotliar}}]{deMedici08}
\bibinfo{author}{\bibfnamefont{L.}~\bibnamefont{de' Medici}},
  \bibinfo{author}{\bibfnamefont{A.}~\bibnamefont{Georges}}, \bibnamefont{and}
  \bibinfo{author}{\bibfnamefont{G.}~\bibnamefont{Kotliar}},
  \bibinfo{journal}{Phys. Rev. B} \textbf{\bibinfo{volume}{77}},
  \bibinfo{pages}{245128} (\bibinfo{year}{2008}).

\bibitem[{\citenamefont{Varma et~al.}(1989)\citenamefont{Varma, Littlewood,
  Schmitt-Rink, Abrahams, and Ruckenstein}}]{Varma89}
\bibinfo{author}{\bibfnamefont{C.~M.} \bibnamefont{Varma}},
  \bibinfo{author}{\bibfnamefont{P.~B.} \bibnamefont{Littlewood}},
  \bibinfo{author}{\bibfnamefont{S.}~\bibnamefont{Schmitt-Rink}},
  \bibinfo{author}{\bibfnamefont{E.}~\bibnamefont{Abrahams}}, \bibnamefont{and}
  \bibinfo{author}{\bibfnamefont{A.~E.} \bibnamefont{Ruckenstein}},
  \bibinfo{journal}{Phys. Rev. Lett.} \textbf{\bibinfo{volume}{63}},
  \bibinfo{pages}{1996} (\bibinfo{year}{1989}).

\bibitem[{\citenamefont{Blanc et~al.}(2009)\citenamefont{Blanc, Gallais,
  Sacuto, Cazayous, M\'easson, Gu, Wen, and Xu}}]{Blanc09}
\bibinfo{author}{\bibfnamefont{S.}~\bibnamefont{Blanc}},
  \bibinfo{author}{\bibfnamefont{Y.}~\bibnamefont{Gallais}},
  \bibinfo{author}{\bibfnamefont{A.}~\bibnamefont{Sacuto}},
  \bibinfo{author}{\bibfnamefont{M.}~\bibnamefont{Cazayous}},
  \bibinfo{author}{\bibfnamefont{M.~A.} \bibnamefont{M\'easson}},
  \bibinfo{author}{\bibfnamefont{G.~D.} \bibnamefont{Gu}},
  \bibinfo{author}{\bibfnamefont{J.~S.} \bibnamefont{Wen}}, \bibnamefont{and}
  \bibinfo{author}{\bibfnamefont{Z.~J.} \bibnamefont{Xu}},
  \bibinfo{journal}{Phys. Rev. B} \textbf{\bibinfo{volume}{80}},
  \bibinfo{pages}{140502} (\bibinfo{year}{2009}).

\bibitem[{\citenamefont{Katsufuji et~al.}(1993)\citenamefont{Katsufuji, Tokura,
  Ido, and Uchida}}]{Katsufuji94}
\bibinfo{author}{\bibfnamefont{T.}~\bibnamefont{Katsufuji}},
  \bibinfo{author}{\bibfnamefont{Y.}~\bibnamefont{Tokura}},
  \bibinfo{author}{\bibfnamefont{T.}~\bibnamefont{Ido}}, \bibnamefont{and}
  \bibinfo{author}{\bibfnamefont{S.}~\bibnamefont{Uchida}},
  \bibinfo{journal}{Phys. Rev. B} \textbf{\bibinfo{volume}{48}},
  \bibinfo{pages}{16131} (\bibinfo{year}{1993}).

\bibitem[{\citenamefont{Venturini et~al.}(2002)\citenamefont{Venturini, Opel,
  Devereaux, Freericks, T\"utt\ifmmode~\mbox{\H{o}}\else \H{o}\fi{}, Revaz,
  Walker, Berger, Forr\'o, and Hackl}}]{Venturini02}
\bibinfo{author}{\bibfnamefont{F.}~\bibnamefont{Venturini}},
  \bibinfo{author}{\bibfnamefont{M.}~\bibnamefont{Opel}},
  \bibinfo{author}{\bibfnamefont{T.~P.} \bibnamefont{Devereaux}},
  \bibinfo{author}{\bibfnamefont{J.~K.} \bibnamefont{Freericks}},
  \bibinfo{author}{\bibfnamefont{I.}~\bibnamefont{T\"utt\ifmmode~\mbox{\H{o}}\%
else \H{o}\fi{}}}, \bibinfo{author}{\bibfnamefont{B.}~\bibnamefont{Revaz}},
  \bibinfo{author}{\bibfnamefont{E.}~\bibnamefont{Walker}},
  \bibinfo{author}{\bibfnamefont{H.}~\bibnamefont{Berger}},
  \bibinfo{author}{\bibfnamefont{L.}~\bibnamefont{Forr\'o}}, \bibnamefont{and}
  \bibinfo{author}{\bibfnamefont{R.}~\bibnamefont{Hackl}},
  \bibinfo{journal}{Phys. Rev. Lett.} \textbf{\bibinfo{volume}{89}},
  \bibinfo{pages}{107003} (\bibinfo{year}{2002}).

\bibitem[{\citenamefont{Hackl et~al.}(2005)\citenamefont{Hackl, Tassini,
  Venturini, Hartinger, Erb, Kikugawa, and Fujita}}]{Hackl05}
\bibinfo{author}{\bibfnamefont{R.}~\bibnamefont{Hackl}},
  \bibinfo{author}{\bibfnamefont{L.}~\bibnamefont{Tassini}},
  \bibinfo{author}{\bibfnamefont{F.}~\bibnamefont{Venturini}},
  \bibinfo{author}{\bibfnamefont{C.}~\bibnamefont{Hartinger}},
  \bibinfo{author}{\bibfnamefont{A.}~\bibnamefont{Erb}},
  \bibinfo{author}{\bibfnamefont{N.}~\bibnamefont{Kikugawa}}, \bibnamefont{and}
  \bibinfo{author}{\bibfnamefont{T.}~\bibnamefont{Fujita}},
  \emph{\bibinfo{title}{Advances in Solid State Physics}}
  (\bibinfo{publisher}{Springer-Verlag, Berlin}, \bibinfo{year}{2005}),
  vol.~\bibinfo{volume}{45}, p. \bibinfo{pages}{227}.

\bibitem[{\citenamefont{Rotter et~al.}(1991)\citenamefont{Rotter, Schlesinger,
  Collins, Holtzberg, Field, Welp, Crabtree, Liu, Fang, Vandervoort
  et~al.}}]{Rotter91}
\bibinfo{author}{\bibfnamefont{L.~D.} \bibnamefont{Rotter}},
  \bibinfo{author}{\bibfnamefont{Z.}~\bibnamefont{Schlesinger}},
  \bibinfo{author}{\bibfnamefont{R.~T.} \bibnamefont{Collins}},
  \bibinfo{author}{\bibfnamefont{F.}~\bibnamefont{Holtzberg}},
  \bibinfo{author}{\bibfnamefont{C.}~\bibnamefont{Field}},
  \bibinfo{author}{\bibfnamefont{U.~W.} \bibnamefont{Welp}},
  \bibinfo{author}{\bibfnamefont{G.~W.} \bibnamefont{Crabtree}},
  \bibinfo{author}{\bibfnamefont{J.~Z.} \bibnamefont{Liu}},
  \bibinfo{author}{\bibfnamefont{Y.}~\bibnamefont{Fang}},
  \bibinfo{author}{\bibfnamefont{K.~G.} \bibnamefont{Vandervoort}},
  \bibnamefont{et~al.}, \bibinfo{journal}{Phys. Rev. Lett.}
  \textbf{\bibinfo{volume}{67}}, \bibinfo{pages}{2741} (\bibinfo{year}{1991}).

\bibitem[{\citenamefont{Jakli$\check{c}$ and
  Prelov$\check{s}$ek}(2000)}]{Jaklic00}
\bibinfo{author}{\bibfnamefont{J.}~\bibnamefont{Jakli$\check{c}$}}
  \bibnamefont{and}
  \bibinfo{author}{\bibfnamefont{P.}~\bibnamefont{Prelov$\check{s}$ek}},
  \bibinfo{journal}{Advances in Physics} \textbf{\bibinfo{volume}{49}},
  \bibinfo{pages}{1} (\bibinfo{year}{2000}).

\bibitem[{\citenamefont{Abanov et~al.}(2001{\natexlab{b}})\citenamefont{Abanov,
  Chubukov, and Schmalian}}]{Abanov01b}
\bibinfo{author}{\bibfnamefont{A.}~\bibnamefont{Abanov}},
  \bibinfo{author}{\bibfnamefont{A.~V.} \bibnamefont{Chubukov}},
  \bibnamefont{and}
  \bibinfo{author}{\bibfnamefont{J.}~\bibnamefont{Schmalian}},
  \bibinfo{journal}{Phys. Rev. B} \textbf{\bibinfo{volume}{63}},
  \bibinfo{pages}{180510} (\bibinfo{year}{2001}{\natexlab{b}}).

\bibitem[{\citenamefont{Schachinger et~al.}(2003)\citenamefont{Schachinger, Tu,
  and Carbotte}}]{Schachinger03}
\bibinfo{author}{\bibfnamefont{E.}~\bibnamefont{Schachinger}},
  \bibinfo{author}{\bibfnamefont{J.~J.} \bibnamefont{Tu}}, \bibnamefont{and}
  \bibinfo{author}{\bibfnamefont{J.~P.} \bibnamefont{Carbotte}},
  \bibinfo{journal}{Phys. Rev. B} \textbf{\bibinfo{volume}{67}},
  \bibinfo{pages}{214508} (\bibinfo{year}{2003}).

\bibitem[{\citenamefont{Haule and Kotliar}(2007)}]{Haule07}
\bibinfo{author}{\bibfnamefont{K.}~\bibnamefont{Haule}} \bibnamefont{and}
  \bibinfo{author}{\bibfnamefont{G.}~\bibnamefont{Kotliar}},
  \bibinfo{journal}{EPL} \textbf{\bibinfo{volume}{77}}, \bibinfo{pages}{27007}
  (\bibinfo{year}{2007}).

\bibitem[{\citenamefont{Chakraborty et~al.}(2008)\citenamefont{Chakraborty,
  Galanakis, and Phillips}}]{Chakraborty08}
\bibinfo{author}{\bibfnamefont{S.}~\bibnamefont{Chakraborty}},
  \bibinfo{author}{\bibfnamefont{D.}~\bibnamefont{Galanakis}},
  \bibnamefont{and} \bibinfo{author}{\bibfnamefont{P.}~\bibnamefont{Phillips}},
  \bibinfo{journal}{Phys. Rev. B} \textbf{\bibinfo{volume}{78}},
  \bibinfo{pages}{212504} (\bibinfo{year}{2008}).

\bibitem[{\citenamefont{Uchida et~al.}(1991)\citenamefont{Uchida, Ido, Takagi,
  Arima, Tokura, and Tajima}}]{Uchida91}
\bibinfo{author}{\bibfnamefont{S.}~\bibnamefont{Uchida}},
  \bibinfo{author}{\bibfnamefont{T.}~\bibnamefont{Ido}},
  \bibinfo{author}{\bibfnamefont{H.}~\bibnamefont{Takagi}},
  \bibinfo{author}{\bibfnamefont{T.}~\bibnamefont{Arima}},
  \bibinfo{author}{\bibfnamefont{Y.}~\bibnamefont{Tokura}}, \bibnamefont{and}
  \bibinfo{author}{\bibfnamefont{S.}~\bibnamefont{Tajima}},
  \bibinfo{journal}{Phys. Rev. B} \textbf{\bibinfo{volume}{43}},
  \bibinfo{pages}{7942} (\bibinfo{year}{1991}).

\bibitem[{\citenamefont{Santander-Syro
  et~al.}(2003)\citenamefont{Santander-Syro, Lobo, Bontemps, Konstantinovic,
  Li, and Raffy}}]{Santander03}
\bibinfo{author}{\bibfnamefont{A.~F.} \bibnamefont{Santander-Syro}},
  \bibinfo{author}{\bibfnamefont{R.~P. S.~M.} \bibnamefont{Lobo}},
  \bibinfo{author}{\bibfnamefont{N.}~\bibnamefont{Bontemps}},
  \bibinfo{author}{\bibfnamefont{Z.}~\bibnamefont{Konstantinovic}},
  \bibinfo{author}{\bibfnamefont{Z.~Z.} \bibnamefont{Li}}, \bibnamefont{and}
  \bibinfo{author}{\bibfnamefont{H.}~\bibnamefont{Raffy}},
  \bibinfo{journal}{EPL} \textbf{\bibinfo{volume}{62}}, \bibinfo{pages}{568}
  (\bibinfo{year}{2003}).

\bibitem[{\citenamefont{Carbone et~al.}(2006)\citenamefont{Carbone, Kuzmenko,
  Molegraaf, van Heumen, Lukovac, Marsiglio, van~der Marel, Haule, Kotliar,
  Berger et~al.}}]{Carbone06}
\bibinfo{author}{\bibfnamefont{F.}~\bibnamefont{Carbone}},
  \bibinfo{author}{\bibfnamefont{A.~B.} \bibnamefont{Kuzmenko}},
  \bibinfo{author}{\bibfnamefont{H.~J.~A.} \bibnamefont{Molegraaf}},
  \bibinfo{author}{\bibfnamefont{E.}~\bibnamefont{van Heumen}},
  \bibinfo{author}{\bibfnamefont{V.}~\bibnamefont{Lukovac}},
  \bibinfo{author}{\bibfnamefont{F.}~\bibnamefont{Marsiglio}},
  \bibinfo{author}{\bibfnamefont{D.}~\bibnamefont{van~der Marel}},
  \bibinfo{author}{\bibfnamefont{K.}~\bibnamefont{Haule}},
  \bibinfo{author}{\bibfnamefont{G.}~\bibnamefont{Kotliar}},
  \bibinfo{author}{\bibfnamefont{H.}~\bibnamefont{Berger}},
  \bibnamefont{et~al.}, \bibinfo{journal}{Phys. Rev. B}
  \textbf{\bibinfo{volume}{74}}, \bibinfo{pages}{064510}
  (\bibinfo{year}{2006}).

\bibitem[{\citenamefont{Millis et~al.}(2005)\citenamefont{Millis, Zimmers,
  Lobo, Bontemps, and Homes}}]{Millis05}
\bibinfo{author}{\bibfnamefont{A.~J.} \bibnamefont{Millis}},
  \bibinfo{author}{\bibfnamefont{A.}~\bibnamefont{Zimmers}},
  \bibinfo{author}{\bibfnamefont{R.~P. S.~M.} \bibnamefont{Lobo}},
  \bibinfo{author}{\bibfnamefont{N.}~\bibnamefont{Bontemps}}, \bibnamefont{and}
  \bibinfo{author}{\bibfnamefont{C.~C.} \bibnamefont{Homes}},
  \bibinfo{journal}{Phys. Rev. B} \textbf{\bibinfo{volume}{72}},
  \bibinfo{pages}{224517} (\bibinfo{year}{2005}).

\bibitem[{\citenamefont{Timusk et~al.}(1995)\citenamefont{Timusk, Homes, and
  Reichardt}}]{Timusk95}
\bibinfo{author}{\bibfnamefont{T.}~\bibnamefont{Timusk}},
  \bibinfo{author}{\bibfnamefont{C.}~\bibnamefont{Homes}}, \bibnamefont{and}
  \bibinfo{author}{\bibfnamefont{W.}~\bibnamefont{Reichardt}},
  \bibinfo{howpublished}{in International Workshop on the Anharmonic Properties
  of High T$_c$ Cuprates, Bled, Slovenia, edited by G. Ruani (World Scientific,
  Singapore), p.121} (\bibinfo{year}{1995}).

\bibitem[{\citenamefont{H\"{u}fner et~al.}(2008)\citenamefont{H\"{u}fner,
  Hossain, Damascelli, and Sawatzky}}]{Hufner08}
\bibinfo{author}{\bibfnamefont{S.}~\bibnamefont{H\"{u}fner}},
  \bibinfo{author}{\bibfnamefont{M.~A.} \bibnamefont{Hossain}},
  \bibinfo{author}{\bibfnamefont{A.}~\bibnamefont{Damascelli}},
  \bibnamefont{and} \bibinfo{author}{\bibfnamefont{G.~A.}
  \bibnamefont{Sawatzky}}, \bibinfo{journal}{Rep. Prog. Phys.}
  \textbf{\bibinfo{volume}{71}}, \bibinfo{pages}{062501}
  (\bibinfo{year}{2008}).

\bibitem[{\citenamefont{Vidhyadhiraja et~al.}(2009)\citenamefont{Vidhyadhiraja,
  Macridin, \ifmmode~\mbox{\c{S}}\else \c{S}\fi{}en, Jarrell, and
  Ma}}]{Vidhyadhiraja09}
\bibinfo{author}{\bibfnamefont{N.~S.} \bibnamefont{Vidhyadhiraja}},
  \bibinfo{author}{\bibfnamefont{A.}~\bibnamefont{Macridin}},
  \bibinfo{author}{\bibfnamefont{C.}~\bibnamefont{\ifmmode~\mbox{\c{S}}\else
  \c{S}\fi{}en}}, \bibinfo{author}{\bibfnamefont{M.}~\bibnamefont{Jarrell}},
  \bibnamefont{and} \bibinfo{author}{\bibfnamefont{M.}~\bibnamefont{Ma}},
  \bibinfo{journal}{Phys. Rev. Lett.} \textbf{\bibinfo{volume}{102}},
  \bibinfo{pages}{206407} (\bibinfo{year}{2009}).

\bibitem[{\citenamefont{Albuquerque et~al.}(2007)\citenamefont{Albuquerque,
  Alet, Corboz et~al.}}]{ALPS}
\bibinfo{author}{\bibfnamefont{A.}~\bibnamefont{Albuquerque}},
  \bibinfo{author}{\bibfnamefont{F.}~\bibnamefont{Alet}},
  \bibinfo{author}{\bibfnamefont{P.}~\bibnamefont{Corboz}},
  \bibnamefont{et~al.}, \bibinfo{journal}{Journal of Magnetism and Magnetic
  Materials} \textbf{\bibinfo{volume}{310}}, \bibinfo{pages}{1187}
  (\bibinfo{year}{2007}).

\end{thebibliography}
\end{document}